\documentclass[prl,aps,superscriptaddress,twocolumn,amssymb,amsmath,longbibliography]{revtex4-2}
\usepackage{graphicx}
\usepackage{theorem}
\usepackage{dcolumn}
\usepackage{bm}
\usepackage{mathrsfs}
\usepackage{setspace}

\begin{document}
\title{Universal anomalous turbulent diffusion in quantum fluids}

\author{Satoshi Yui}
\thanks{Equal contributions}
\affiliation{Research and Education Center for Natural Sciences, Keio University, 4-1-1 Hiyoshi, Kohoku-ku, Yokohama 223-8521, Japan}

\author{Yuan Tang}
\thanks{Equal contributions}
\affiliation{National High Magnetic Field Laboratory, 1800 East Paul Dirac Drive, Tallahassee, Florida 32310, USA}
\affiliation{Mechanical Engineering Department, FAMU-FSU College of Engineering, Florida State University, Tallahassee, Florida 32310, USA}

\author{Wei Guo}
\email[Email: ]{wguo@magnet.fsu.edu}
\affiliation{National High Magnetic Field Laboratory, 1800 East Paul Dirac Drive, Tallahassee, Florida 32310, USA}
\affiliation{Mechanical Engineering Department, FAMU-FSU College of Engineering, Florida State University, Tallahassee, Florida 32310, USA}

\author{Hiromichi Kobayashi}
\email[Email: ]{hkobayas@keio.jp}
\affiliation{Research and Education Center for Natural Sciences, Keio University, 4-1-1 Hiyoshi, Kohoku-ku, Yokohama 223-8521, Japan}
\affiliation{Department of Physics, Keio University, 4-1-1 Hiyoshi, Kohoku-ku, Yokohama 223-8521, Japan}

\author{Makoto Tsubota}
\email[Email: ]{tsubota@osaka-cu.ac.jp}
\affiliation{Department of Physics \& Nambu Yoichiro Institute of Theoretical and Experimental Physics (NITEP) \& The OCU Advanced Research Institute for Natural Science and Technology (OCARINA), Osaka City University, 3-3-138 Sugimoto, Sumiyoshi-ku, Osaka 558-8585, Japan}

\date{\today}

\begin{abstract}
In classical viscous fluids, turbulent eddies are known to be responsible for the rapid spreading of embedded particles. But in an inviscid quantum fluid where the turbulence is induced by a chaotic tangle of quantized vortices, dispersion of the particles is achieved via a non-classical mechanism, i.e., their binding to the evolving quantized vortices. However, there is limited existing knowledge on how the vortices diffuse and spread in turbulent quantum fluids. Here we report a systematic numerical study of the apparent diffusion of vortices in a random vortex tangle in superfluid helium-4 using full Biot-Savart simulation. We reveal that the vortices in pure superfluid exhibit a universal anomalous diffusion (superdiffusion) at small times, which transits to normal diffusion at large times. This behavior is found to be caused by a generic scaling property of the vortex velocity, which should exist in all quantum fluids where the Biot-Savart law governs the vortex motion. Our simulation at finite temperatures also nicely reproduces recent experimental observations. The knowledge obtained from this study may form the foundation for understanding turbulent transport and universal vortex dynamics in various condensed-matter and cosmic quantum fluids.
\end{abstract}
\maketitle

Turbulent diffusion in classical fluids has been studied extensively due to its wide range of applications such as chemical mixing in star formation~\cite{Feng-2014-Nature}, pollution migration~\cite{Petaja-2016-SR}, and airborne virus transmission~\cite{Wang-2021-Science}. It has been known that the turbulent eddies can carry embedded particles and facilitate their spreading~\cite{Sreeni-2019-PNAS}. However, this knowledge does not apply to inviscid quantum fluids such as superfluid helium-4 (He II) and atomic Bose-Einstein condensates (BECs), since the injected particles are not entrained by the superfluid flow at low temperatures. Instead, a distinct transport mechanism exists.

In a quantum fluid, turbulence can be induced by a chaotic tangle of quantized vortex lines~\cite{Vinen-2002-JLP}, which are line-shaped topological defects featured by a circulating flow with a fixed circulation $\kappa=h/m$, where $h$ is Plank's constant and $m$ is the mass of the bosons constituting the superfluid~\cite{Tilley-1990-book}. The vortices evolve with time chaotically, and they also undergo reconnections when they move across each other~\cite{Koplik-1993-PRL}. Impurity particles in the quantum fluid can bind to the vortex cores and subsequently move together with the vortices~\cite{Gomez-2014-Science, Bewley-2006-Nature, Zmeev-2013-PRL,Mastracci-2019-PRF}. Knowing how vortices diffuse in space is crucial for understanding turbulent transport in quantum fluids. Such knowledge could potentially benefit various studies such as chemical dopant behaviors in superfluid nanodroplets~\cite{Toennies-2001-PT}, vortex-aided nanowire fabrication~\cite{Latimer-2014-NL}, and possibly matter aggregation around cosmic strings~\cite{Silk-1984-PRL}.

So far, there have been very limited studies on the apparent diffusion of vortices in a quantum-fluid turbulence. On the theoretical side, the overall expansion of a decaying random vortex tangle near a solid wall~\cite{Tsubota-2003-PB} and in bulk He II~\cite{Rickinson-2019-PRB} was simulated. But these studies only provide limited insights into the diffusion behavior of individual vortices in a fully developed tangle. In a recent experiment, Tang \emph{et al.} decorated a random tangle of vortices in He II with solidified deuterium tracer particles and observed that the vortices undergo anomalous diffusion (superdiffusion) at small times~\cite{Tang-2021-PNAS}. Their measured diffusion time exponent appears to be insensitive to variations in both the temperature and the vortex-line density, suggesting possible generic nature of this vortex-diffusion behavior. However, since the experiment was conducted only in a narrow temperature range (i.e., 1.7 K to 2.0 K), it is not clear whether the observed phenomenon holds at other temperatures and whether it is truly generic for all quantum fluids.

\begin{figure*}[t]
\centering
\includegraphics[width=1\linewidth]{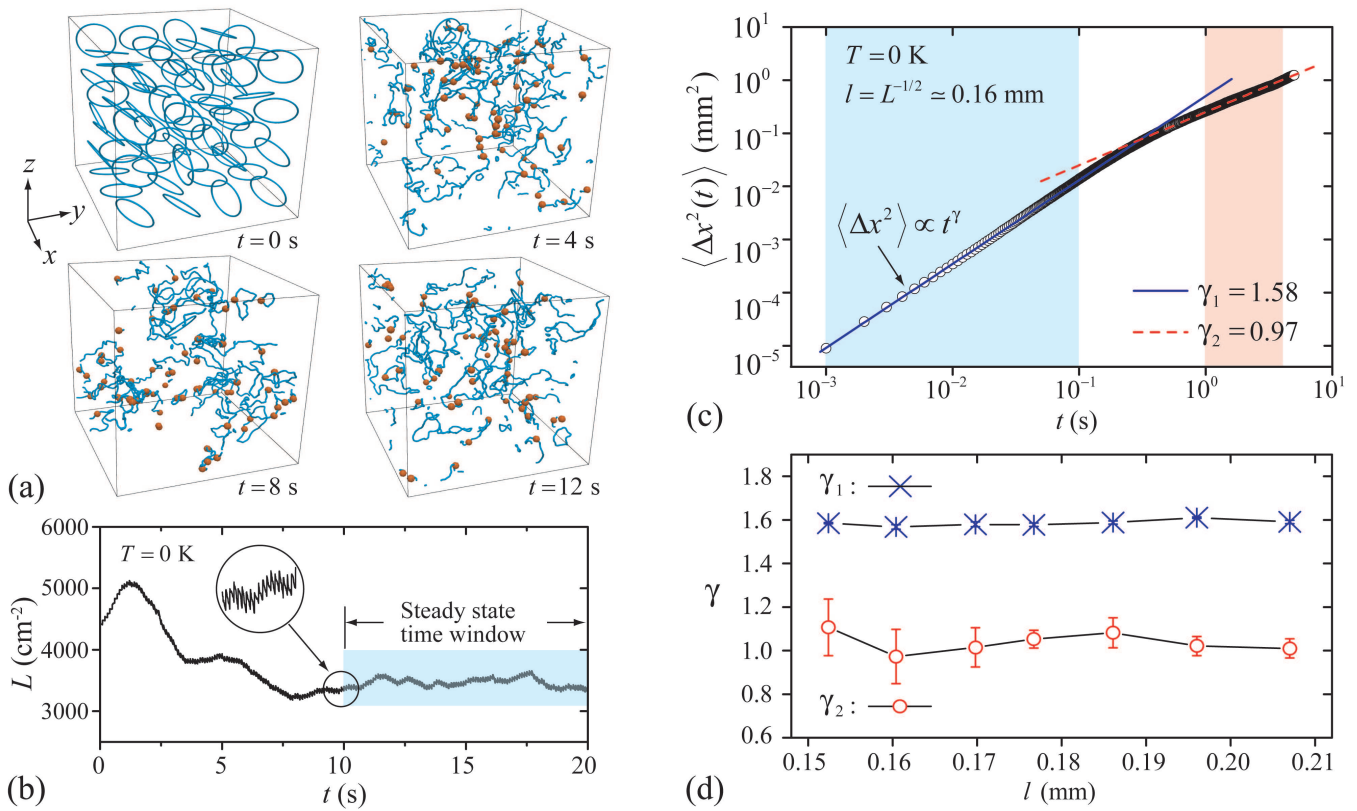}
\caption{(a) Snapshots of an evolving random vortex tangle with $t_{in}=0.08$ s. The red dots represents the vortex-filament points tracked for diffusion analysis. (b) Variation of the corresponding vortex-line density $L(t)$. The blow-up in the inset shows fluctuations in $L$ due to ring injections. (c) Representative data showing the mean square displacement of the vortices $\langle\Delta x^2(t)\rangle$ in the $x$ direction. The solid and dashed lines are power-law fits to the data in the shaded regions. (d) Diffusion exponents $\gamma$ averaged over $x$, $y$, and $z$ directions versus the mean vortex-line spacing $\ell=L^{-1/2}$. The error bars (barely visible for $\gamma_1$) represent the standard deviations of the $\gamma$ values for diffusion in the three axial directions.}
\label{Fig1}
\end{figure*}

In this paper, we report a systematic numerical study of the apparent diffusion of individual vortices in a random vortex tangle in He II using the full Biot-Savart simulation~\cite{Adachi-2010-PRB}. A random tangle does not generate any large-scale flows to advect the vortices~\cite{Volovik-2003-JETP,Walmsley-2008-PRL}, which makes it feasible to observe intrinsic vortex diffusion behaviors that are characteristics of the Biot-Savart law. We reveal that in pure superfluid the vortices indeed undergo superdiffusion at small times with a \emph{universal} diffusion exponent regardless how dense the tangle is. At large times, the superdiffusion transits to normal diffusion due to vortex reconnections. Our analysis shows that this universal diffusion behavior is caused by a generic temporal correlation of the vortex velocity, which should exist in all quantum fluids where the Biot-Savart law applies. At finite temperatures, the viscous effect is found to only mildly affect the vortex diffusion, which nicely explains the experimental observations.

Note that random vortex tangles are indeed ubiquitous: they can be produced by counterflow in quantum two-fluid systems such as He II~\cite{Vinen-1957-PRS-I,Gao-2016-JETP}, atomic BECs~\cite{Takeuchi-2010-PRL}, superfluid neutron stars~\cite{Greenstein-1970-Nature,Haskell-2020-MNRAS}, and dark matter BECs~\cite{Hartman-2021-AA,Sikivie-2009-PRL}; and they can also spontaneously emerge following a second-order phase transition in quantum fluids via the Kibble-Zurek mechanism~\cite{Zurek-1985-Nature,Stagg-2016-PRA}. Therefore, the knowledge obtained in our study may also offer valuable insights into the evolution and quenching dynamics of these diverse quantum-fluid systems.

\emph{Vortex diffusion in a pure superfluid.}--Like many other quantum fluids, He II can be considered as a mixture of two miscible fluid components, i.e., an inviscid superfluid (the condensate) and a viscous normal fluid (collection of thermal quasiparticles)~\cite{Landau-book}. The normal-fluid fraction in He II drops with decreasing temperature and becomes negligible below 1 K. Experimentally, a random vortex tangle can be produced in pure superfluid helium at zero-temperature limit by injecting small vortex rings~\cite{Walmsley-2008-PRL}. Here, we adopt a similar method numerically to study vortex diffusion in pure superfluid. As shown in Fig.~\ref{Fig1}~(a), we first place 64 randomly oriented vortex rings (radius: $R_{in}=0.11$ mm) in a cubical computational box (side length: $D=1$ mm) with periodic boundary conditions in all three axial directions. These vortice are described by the vortex filament model~\cite{Schwarz-1988-PRB}, and each vortex filament is discretized into a series of points. In the absence of the normal fluid, a vortex-filament point at $\mathbf{s}$ moves at the local superfluid velocity $\mathbf{v}_s(\mathbf{s})$ induced by all the vortices as given by the Boit-Savart law~\cite{Schwarz-1988-PRB,Tsubota-2000-PRB}:
\begin{equation}
\frac{d\mathbf{s}}{dt}=\mathbf{v}_s(\mathbf{s})=\frac{\kappa}{4\pi}\int\frac{(\mathbf{s_1}-\mathbf{s})\times d\mathbf{s_1}}{|\mathbf{s_1}-\mathbf{s}|^3}.
\label{Eq1}
\end{equation}
The time evolution of the vortices can be obtained through a temporal integration of equation~(\ref{Eq1}) (see details and movies in Supplemental Materials). When two vortex filaments approach to have a minimum separation less than 0.008 mm, we reconnect them at the location of the minimum separation following the procedures as detailed in references~\cite{Tsubota-2000-PRB,Baggaley-2012-JLTP}. We also inject a randomly oriented vortex ring of radius $R_{in}$ in the computational box with a repetition time $t_{in}$ to balance the cascade loss of the vortices~\cite{Tsubota-2000-PRB}. Fig.~\ref{Fig1}~(a) shows the evolution of the vortex tangle with $t_{in}=0.08$ s. The variation of the vortex-line density $L$ (i.e., length of the vortices per unit volume) is shown in Fig.~\ref{Fig1}~(b). It is clear that after about 10 s, $L$ settles to a nearly constant level. This steady-state $L$ level can be tuned by varying $t_{in}$.

To study vortex diffusion, we track randomly chosen vortex-filament points and analyze their mean square displacement (MSD) along each axis (see Supplemental Materials). Fig.~\ref{Fig1}~(c) shows the MSD of the vortices in the $x$ direction $\langle\Delta x^2(t)\rangle=\langle[x(t_0+t)-x(t_0)]^2\rangle$ in a representative case, where the angle brackets denote an ensemble average of all the tracked vortex-filament points in the steady-state time window. Usually, a power-law scaling $\langle\Delta x^2(t)\rangle\propto t^\gamma$ is expected, where the exponent $\gamma$ defines different diffusion regimes, i.e., normal diffusion ($\gamma=1$) and anomalous diffusion (superdiffusion at $\gamma>$1 and subdiffusion at $\gamma<$1)~\cite{Ben-2000-book}. Our data exhibit a clear superdiffusion regime ($\gamma_1=1.58$) at small $t$ and a normal diffusion regime ($\gamma_2=0.97$) at large $t$.

Simulations conducted at other $L$ also show similar behaviors. The derived $\gamma_1$ and $\gamma_2$ are plotted in Fig.~\ref{Fig1}~(d) as a function of the mean vortex-line spacing $\ell=L^{-1/2}$. It is clear that $\gamma_1\simeq1.6$ is universal regardless how dense the tangle is. $\gamma_2$ is around 1 but has sizable variations in the three axial directions. These variations are caused by the reduced sample numbers at large $t$ as well as the non-uniformity of the tangle at large length scales due to the limited size of the computational box (see Supplemental Materials for details).

\begin{figure}[t]
\centering
\includegraphics[width=1\linewidth]{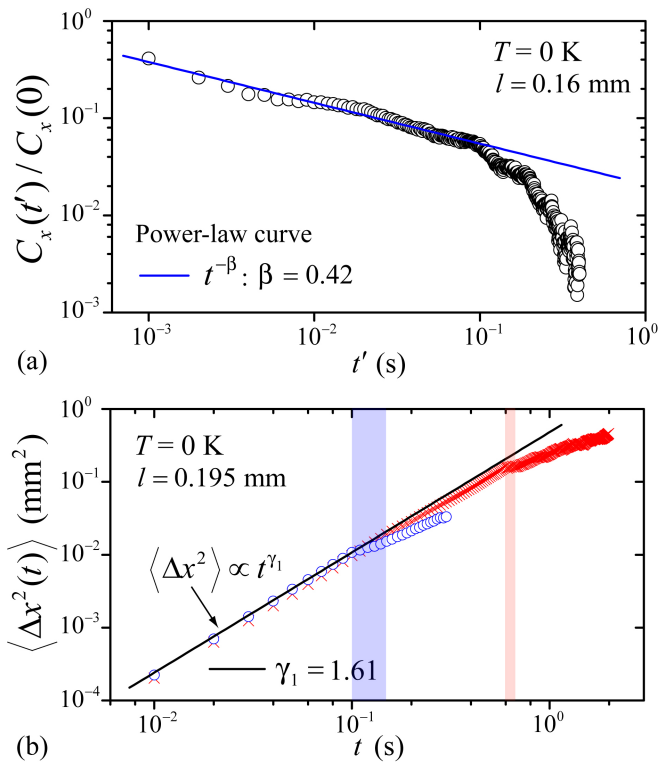}
\caption{(a) Vortex-velocity temporal correlation function $C_x(t')$ for a representative tangle with $\ell=0.16$ mm exhibiting a power-law scaling that leads to the observed superdiffusion. (b) $\langle\Delta x^2(t)\rangle$ data of two selected groups of vortex-filament points (i.e., blue circles and red crosses) whose first reconnection event occurred in the shaded time intervals of the respective colors.
}
\label{Fig2}
\end{figure}

\emph{Mechanism of the universal vortex diffusion.}--To explain the physical mechanism of the observed vortex diffusion behavior, we first note that superdiffusion has been observed in various physical and non-physical systems~\cite{Solomon-1993-PRL,Harris-2012-Nature,Sagi-2012-PRL,Raichlen-2014-PNAS}. For systems involving random walkers, the appearance of superdiffusion is usually attributed to the so-called L\'{e}vy flights, i.e., occasional long-distance hops of the random walkers~\cite{Bouchaud-1990-PR}. These flights lead to a power-law tail of the walker's displacement distribution $P(\Delta x)\propto|\Delta x|^{-\alpha}$, which is flat enough (i.e., $\alpha<3$) to cause superdiffusion~\cite{Bouchaud-1990-PR}. In He II, large displacements of the vortex-filament points over a short time can occur at the locations where the vortices reconnect~\cite{Bewley-2008-PNAS,Paoletti-2008-PRL}. However, we find that these reconnections always result in a tail of the vortex displacement distribution $P(\Delta x)$ steeper than $|\Delta x|^{-3}$ (see Supplemental Materials). Therefore, they cannot account for the observed vortex superdiffusion. On the other hand, superdiffusion may emerge if the motion of the walkers is not completely random but has extended temporal correlations~\cite{Bouchaud-1990-PR,Davison-1989-PRS}. To see this, we write the MSD of a vortex-filament point $\langle\Delta x^2(t)\rangle$ in terms of its velocity $v_x(t)$ as~\cite{Mazzitelli-2004-NJP}:
\begin{equation}
\langle\Delta x^2(t)\rangle=2\int_0^tdt_0\int_0^{t-t_0}dt'\langle v_x(t_0)v_x(t_0+t')\rangle.
\label{Eq2}
\end{equation}
For a fully developed random tangle, the vortex-velocity temporal correlation function $C_x(t',t_0)=\langle v_x(t_0)v_x(t_0+t')\rangle$ only depends on the lapse time $t'$, i.e., $C_x(t',t_0)=C_x(t')$. If $C_x(t')$ shows a power-law scaling $C_x(t')\propto t'^{-\beta}$ over a large time interval, $\langle\Delta x^2(t)\rangle$ then scales as $\langle\Delta x^2(t)\rangle\propto t^{2-\beta}$ in the same time interval and can exhibit superdiffusion when $\beta<1$. In Fig.~\ref{Fig2}~(a), we show the calculated $C_x(t')$ for a representative tangle with $\ell=0.16$ mm. There is a clear power-law scaling with a fitted exponent $\beta=0.42$, which leads to $\langle\Delta x^2(t)\rangle\propto t^{1.58}$, matching nicely the superdiffusion exponent reported in Fig.~\ref{Fig1}. Similar results are obtained for other cases at different $\ell$, which reveals that the universal vortex superdiffusion at small $t$ is caused by the temporal correlation of the vortex velocity. This correlation should be an intrinsic property of the Biot-Savart law applicable to all superfluids.

The transition to the normal diffusion at large $t$ was also observed experimentally by Tang \emph{et al.}~\cite{Tang-2021-PNAS}. They proposed that this transition is caused by vortex reconnections, which effectively randomize the motion of the participating vortex-filament points. This randomization suppresses the velocity temporal correlation and results in the normal diffusion at large $t$ according to equation~(\ref{Eq2}). To verify this view, we analyze the MSD of two representative groups of vortex-filament points whose first reconnection event occurred in the diffusion time interval $0.1-0.15$~s and $0.6-0.65$~s, respectively. Relevant data for the case with $\ell=0.195$ mm are shown in Fig.~\ref{Fig2}~(b). Obvious deviation from the superdiffusion scaling is observed for each group only after their first reconnection event. This observation clearly proves the causality between vortex reconnections and the transition towards the normal diffusion.

\begin{figure}[t]
\centering
\includegraphics[width=1\linewidth]{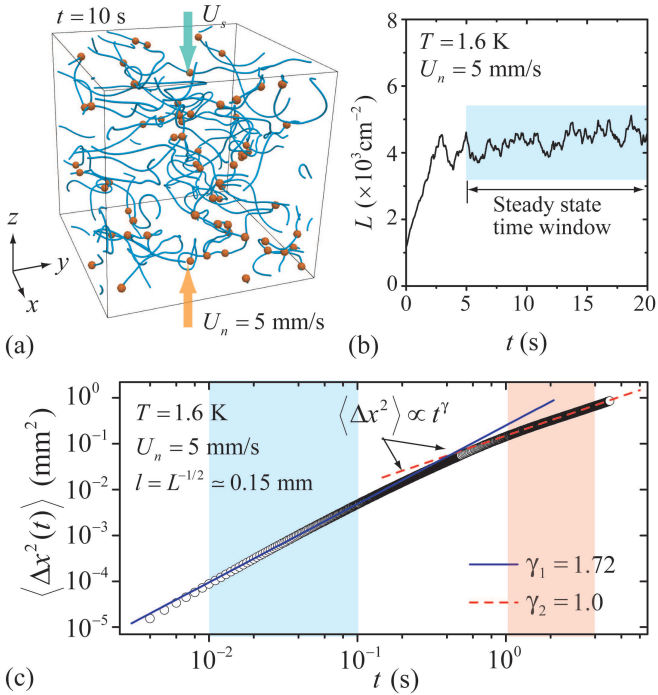}
\caption{(a) A representative snapshot of the vortex tangle produced by counterflow at $T=1.6$ K with $U_n=5$ mm/s. The red dots represent the tracked vortex-filament points. (b) A representative vortex-line density $L(t)$ curve obtained in counterflow. (c) MSD of the vortices $\langle\Delta x^2(t)\rangle$ in the $x$ direction in counterflow. The solid and dashed lines are power-law fits to the data in the shaded regions.
}
\label{Fig3}
\end{figure}

\emph{Finite-temperature effect.}--At finite temperatures, the vortices experience a drag force as they move through the normal fluid due to scattering of the thermal quasiparticles in He II~\cite{Vinen-1957-PRS-III}. The velocity of a vortex-filament point at $\mathbf{s}$ is now given by~\cite{Schwarz-1988-PRB,Tsubota-2000-PRB}:
\begin{equation}
d\mathbf{s}/dt=\mathbf{v}_s(\mathbf{s})+\alpha\mathbf{s}'\times(\mathbf{v}_n-\mathbf{v}_s)-\alpha'\mathbf{s}'\times[\mathbf{s}'\times(\mathbf{v}_n-\mathbf{v}_s)],
\label{Eq3}
\end{equation}
where $\alpha$ and $\alpha'$ are temperature dependant mutual friction coefficients for He II~\cite{Donnelly-1991-B}, $\mathbf{s}'$ is the unit tangent vector along the filament, and $\mathbf{v}_n$ is the normal-fluid velocity. We then generate a steady-state vortex tangle using two distinct methods. The first method is similar to the one adopted at 0 K, i.e., by injecting small vortex rings in the computational box with static normal fluid (i.e., $\mathbf{v}_n=0$). The second method is via thermal counterflow as adopted in the experiment conducted by Tang \emph{et al}~\cite{Tang-2021-PNAS}.

\begin{figure}[t]
\centering
\includegraphics[width=1\linewidth]{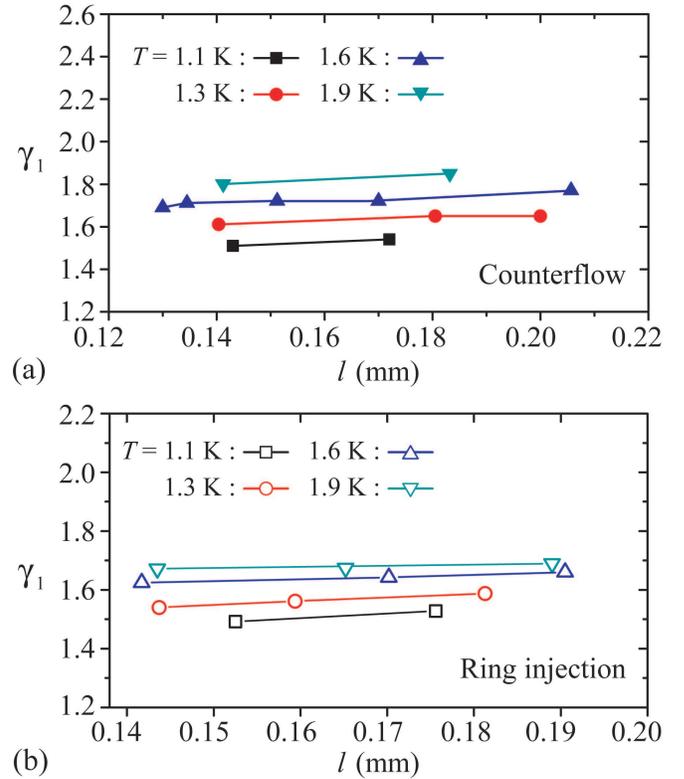}
\caption{(a) Diffusion exponent $\gamma_1$ versus $\ell$ for vortex tangles produced by counterflow at various $T$. (e) $\gamma_1$ versus $\ell$ for tangles produced by vortex-ring injection.
}
\label{Fig4}
\end{figure}

In He II, a counterflow can be generated by an applied heat flux $q$, where the normal fluid moves in the heat flow direction at a mean velocity $U_n=q/\rho sT$ while the superfluid moves oppositely at $U_s=(\rho_n/\rho_s)U_n$~\cite{Landau-book}. Here, $\rho=\rho_n+\rho_s$ is the total density, and $s$ is the He II specific entropy~\cite{Donnelly-1998-JPCRD}. To compare with the experiment where the normal-fluid flow is laminar, we set $\mathbf{v}_n=U_n\mathbf{\hat{e}}_z$ and $\mathbf{v}_s$ as the sum of $-U_s\mathbf{\hat{e}}_z$ and the induced velocity given in equation~(\ref{Eq1}) (see Supplemental Materials). We then place a few randomly oriented seed vortex rings in the computational box. These rings can grow and reconnect, eventually leading to the formation of a steady vortex tangle~\cite{Adachi-2010-PRB}. A snapshot of such a tangle at $T=1.6$~K and $U_n=5$~mm/s is shown in Fig.~\ref{Fig3}~(a), and the line-density evolution is given in Fig.~\ref{Fig3}~(b). In the steady-state time window (i.e., $5-20$~s), we track randomly selected vortex-filament points and analyze their MSD in the directions perpendicular to the counterflow. Representative data for $\langle\Delta x^2(t)\rangle$ are shown in Fig.~\ref{Fig3}~(c). Again, a superdiffusion regime is observed at small $t$, which transits to normal diffusion at large $t$.

In Fig.~\ref{Fig4}~(a) and (b), we collect the derived superdiffusion exponent $\gamma_1$ for tangles generated respectively by counterflow and ring injections at various $\ell$ and $T$. It is clear that $\gamma_1$ is around 1.6 and is nearly independent of $\ell$, which is in good agreement with the experimental observations~\cite{Tang-2021-PNAS}. We also see that $\gamma_1$ increases by less than 0.3 from 1.1~K to 1.9~K, which makes it hardly resolvable in the narrow temperature range examined in the experiment. At a given $T$, the $\gamma_1$ value for tangles produced by counterflow is slightly larger than that for random tangles produced by ring injections. This difference becomes more visible as $T$ increases. Interestingly, it has been known that the vortex tangle produced by counterflow becomes increasingly anisotropic as $T$ increases~\cite{Adachi-2010-PRB}. This tangle anisotropy could be the origin of the observed difference, which is a topic for future research.

\begin{figure}[t]
\centering
\includegraphics[width=1\linewidth]{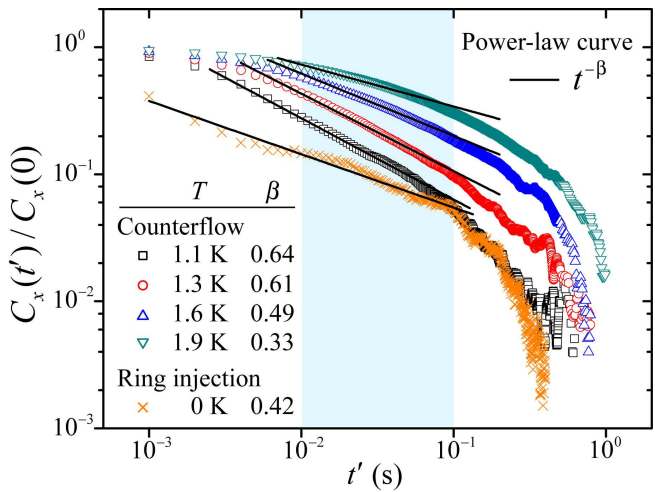}
\caption{Variation of the vortex-velocity temporal correlation function $C_x(t')$ with $T$. The data shown are for tangles with $\ell$ of about $0.14-0.16$ mm. The solid lines represent power-law fits to the data in the shaded region.}
\label{Fig5}
\end{figure}

\emph{Discussion.}--The weak temperature dependence of $\gamma_1$ indeed involves interesting physics. To explain it, we again refer to the vortex-velocity temporal correlation $C_x(t')$. In Fig.~\ref{Fig5}, we show the normalized $C_x(t')/C_x(0)$ curves for vortex tangles produced by counterflow at various $T$ with $\ell$ in the range of $0.14-0.16$ mm. The $C_x(t')/C_x(0)$ curve for the random tangle produced by ring injections at 0 K with $\ell=0.16$ mm is also included as a reference. Compared to the $T=0$ curve, a major difference of the curves at finite $T$ is that they saturate to 1 at larger $t'$. This saturation, which corresponds to ballistic motion of the vortices (i.e., $\langle\Delta x^2(t)\rangle\propto t^2$) according to equation~\ref{Eq2}, is controlled by how the turbulent energy decays. At 0 K, the turbulent energy can cascade down to scales smaller than $\ell$ by exciting Kelvin waves on the vortex lines~\cite{Kivotides-2001-PRL}. These waves result in small-scale deformations and kinks on the vortices (see movies in the Supplemental Materials), which can cause rapid turning of the vortex-velocity vector~\cite{Tsubota-2000-PRB}. This process suppresses the velocity temporal correlation and is the reason why $C_x(t')$ at 0 K remains far from saturation even down to $t'=10^{-3}$~s. At finite $T$, the mutual friction from the normal fluid damps out the Kelvin waves and terminates the energy cascade at scales comparable to $\ell$~\cite{Gao-2016-PRB,Gao-2018-PRB}. Therefore, small-scale deformations on the vortices are all smoothed out, which helps extending the ballistic motion of the vortices to larger $t'$. This extension first makes the $C_x(t')/C_x(0)$ curve steeper in the power-law region, thereby leading to a larger exponent $\beta$ as compared to the 0~K case. As $T$ further increases, the power-law region gradually shrinks and levels off, which reduces $\beta$. Although $\beta$ and $\gamma_1$ are no longer correlated exactly as $\gamma_1=2-\beta$ due to the presence of the saturation regime, the decrease in $\beta$ as $T$ increases still causes $\gamma_1$ to gradually rise as observed.

In summary, our numerical study has produced a comprehensive understanding of the intriguing diffusion behaviors of the quantized vortices in a random vortex tangle. This study may stimulate future research on hidden universal scaling laws in quantum-fluid turbulence pertinent to trajectory statistics of individual vortices.

\begin{acknowledgments}
Y.T. and W.G. are supported by the National Science Foundation under Grant DMR-2100790 and the US Department of Energy under Grant DE-SC0020113. They also acknowledge the support and resources provided by the National High Magnetic Field Laboratory at Florida State University, which is supported by the National Science Foundation Cooperative Agreement No. DMR-1644779 and the state of Florida. S.Y. is supported by the Grant-in-Aid for JSPS Fellow program under Grant No. JP19J00967. M.T. acknowledges the support by the JSPS KAKENHI program under Grant No. JP20H01855.
\end{acknowledgments}

\bibliography{PRL-Diffusion}

\begin{thebibliography}{49}%
\makeatletter
\providecommand \@ifxundefined [1]{%
 \@ifx{#1\undefined}
}%
\providecommand \@ifnum [1]{%
 \ifnum #1\expandafter \@firstoftwo
 \else \expandafter \@secondoftwo
 \fi
}%
\providecommand \@ifx [1]{%
 \ifx #1\expandafter \@firstoftwo
 \else \expandafter \@secondoftwo
 \fi
}%
\providecommand \natexlab [1]{#1}%
\providecommand \enquote  [1]{``#1''}%
\providecommand \bibnamefont  [1]{#1}%
\providecommand \bibfnamefont [1]{#1}%
\providecommand \citenamefont [1]{#1}%
\providecommand \href@noop [0]{\@secondoftwo}%
\providecommand \href [0]{\begingroup \@sanitize@url \@href}%
\providecommand \@href[1]{\@@startlink{#1}\@@href}%
\providecommand \@@href[1]{\endgroup#1\@@endlink}%
\providecommand \@sanitize@url [0]{\catcode `\\12\catcode `\$12\catcode
  `\&12\catcode `\#12\catcode `\^12\catcode `\_12\catcode `\%12\relax}%
\providecommand \@@startlink[1]{}%
\providecommand \@@endlink[0]{}%
\providecommand \url  [0]{\begingroup\@sanitize@url \@url }%
\providecommand \@url [1]{\endgroup\@href {#1}{\urlprefix }}%
\providecommand \urlprefix  [0]{URL }%
\providecommand \Eprint [0]{\href }%
\providecommand \doibase [0]{https://doi.org/}%
\providecommand \selectlanguage [0]{\@gobble}%
\providecommand \bibinfo  [0]{\@secondoftwo}%
\providecommand \bibfield  [0]{\@secondoftwo}%
\providecommand \translation [1]{[#1]}%
\providecommand \BibitemOpen [0]{}%
\providecommand \bibitemStop [0]{}%
\providecommand \bibitemNoStop [0]{.\EOS\space}%
\providecommand \EOS [0]{\spacefactor3000\relax}%
\providecommand \BibitemShut  [1]{\csname bibitem#1\endcsname}%
\let\auto@bib@innerbib\@empty
\bibitem [{\citenamefont {Feng}\ and\ \citenamefont
  {Krumholz}(2014)}]{Feng-2014-Nature}%
  \BibitemOpen
  \bibfield  {author} {\bibinfo {author} {\bibfnamefont {Y.}~\bibnamefont
  {Feng}}\ and\ \bibinfo {author} {\bibfnamefont {M.}~\bibnamefont
  {Krumholz}},\ }\bibfield  {title} {\bibinfo {title} {Early turbulent mixing
  as the origin of chemical homogeneity in open star clusters},\ }\href
  {https://doi.org/10.1038/nature13662} {\bibfield  {journal} {\bibinfo
  {journal} {Nature}\ }\textbf {\bibinfo {volume} {513}},\ \bibinfo {pages}
  {523} (\bibinfo {year} {2014})}\BibitemShut {NoStop}%
\bibitem [{\citenamefont
  {Pet$\ddot{\mathrm{a}}$j$\ddot{\mathrm{a}}$~\emph{et~al.}}(2016)}]{Petaja-2016-SR}%
  \BibitemOpen
  \bibfield  {author} {\bibinfo {author} {\bibfnamefont {T.}~\bibnamefont
  {Pet$\ddot{\mathrm{a}}$j$\ddot{\mathrm{a}}$~\emph{et~al.}}},\ }\bibfield
  {title} {\bibinfo {title} {Enhanced air pollution via aerosol-boundary layer
  feedback in china},\ }\href {https://doi.org/10.1038/srep18998} {\bibfield
  {journal} {\bibinfo  {journal} {Sci. Rep.}\ }\textbf {\bibinfo {volume}
  {6}},\ \bibinfo {pages} {18998} (\bibinfo {year} {2016})}\BibitemShut
  {NoStop}%
\bibitem [{\citenamefont {Wang}\ \emph {et~al.}(2021)\citenamefont {Wang},
  \citenamefont {Prather}, \citenamefont {Sznitman}, \citenamefont {Jimenez},
  \citenamefont {Lakdawala}, \citenamefont {Tufekci},\ and\ \citenamefont
  {Marr}}]{Wang-2021-Science}%
  \BibitemOpen
  \bibfield  {author} {\bibinfo {author} {\bibfnamefont {C.~C.}\ \bibnamefont
  {Wang}}, \bibinfo {author} {\bibfnamefont {K.~A.}\ \bibnamefont {Prather}},
  \bibinfo {author} {\bibfnamefont {J.}~\bibnamefont {Sznitman}}, \bibinfo
  {author} {\bibfnamefont {J.~L.}\ \bibnamefont {Jimenez}}, \bibinfo {author}
  {\bibfnamefont {S.~S.}\ \bibnamefont {Lakdawala}}, \bibinfo {author}
  {\bibfnamefont {Z.}~\bibnamefont {Tufekci}},\ and\ \bibinfo {author}
  {\bibfnamefont {L.~C.}\ \bibnamefont {Marr}},\ }\bibfield  {title} {\bibinfo
  {title} {Airborne transmission of respiratory viruses},\ }\href
  {https://doi.org/10.1126/science.abd9149} {\bibfield  {journal} {\bibinfo
  {journal} {Science}\ }\textbf {\bibinfo {volume} {373}},\ \bibinfo {pages}
  {eabd9149} (\bibinfo {year} {2021})}\BibitemShut {NoStop}%
\bibitem [{\citenamefont {Sreenivasan}(2019)}]{Sreeni-2019-PNAS}%
  \BibitemOpen
  \bibfield  {author} {\bibinfo {author} {\bibfnamefont {K.~R.}\ \bibnamefont
  {Sreenivasan}},\ }\bibfield  {title} {\bibinfo {title} {Turbulent mixing: A
  perspective},\ }\href {https://doi.org/10.1073/pnas.1800463115} {\bibfield
  {journal} {\bibinfo  {journal} {Proc. Natl. Acad. Sci. U.S.A}\ }\textbf
  {\bibinfo {volume} {116}},\ \bibinfo {pages} {18175} (\bibinfo {year}
  {2019})}\BibitemShut {NoStop}%
\bibitem [{\citenamefont {Vinen}\ and\ \citenamefont
  {Niemela}(2002)}]{Vinen-2002-JLP}%
  \BibitemOpen
  \bibfield  {author} {\bibinfo {author} {\bibfnamefont {W.}~\bibnamefont
  {Vinen}}\ and\ \bibinfo {author} {\bibfnamefont {J.}~\bibnamefont
  {Niemela}},\ }\bibfield  {title} {\bibinfo {title} {Quantum turbulence},\
  }\href@noop {} {\bibfield  {journal} {\bibinfo  {journal} {J. Low Temp.
  Phys.}\ }\textbf {\bibinfo {volume} {128}},\ \bibinfo {pages} {167} (\bibinfo
  {year} {2002})}\BibitemShut {NoStop}%
\bibitem [{\citenamefont {Tilley}\ and\ \citenamefont
  {Tilley}(1990)}]{Tilley-1990-book}%
  \BibitemOpen
  \bibfield  {author} {\bibinfo {author} {\bibfnamefont {D.}~\bibnamefont
  {Tilley}}\ and\ \bibinfo {author} {\bibfnamefont {J.}~\bibnamefont
  {Tilley}},\ }\href@noop {} {\emph {\bibinfo {title} {Superfluidity and
  Superconductivity}}},\ \bibinfo {edition} {3rd}\ ed.\ (\bibinfo  {publisher}
  {Institute of Physics},\ \bibinfo {address} {Bristol},\ \bibinfo {year}
  {1990})\BibitemShut {NoStop}%
\bibitem [{\citenamefont {Koplik}\ and\ \citenamefont
  {Levine}(1993)}]{Koplik-1993-PRL}%
  \BibitemOpen
  \bibfield  {author} {\bibinfo {author} {\bibfnamefont {J.}~\bibnamefont
  {Koplik}}\ and\ \bibinfo {author} {\bibfnamefont {H.}~\bibnamefont
  {Levine}},\ }\bibfield  {title} {\bibinfo {title} {Vortex reconnection in
  superfluid helium},\ }\href {https://doi.org/10.1103/PhysRevLett.71.1375}
  {\bibfield  {journal} {\bibinfo  {journal} {Phys. Rev. Lett.}\ }\textbf
  {\bibinfo {volume} {71}},\ \bibinfo {pages} {1375} (\bibinfo {year}
  {1993})}\BibitemShut {NoStop}%
\bibitem [{\citenamefont {Gomez~$\emph{et al.}$}(2014)}]{Gomez-2014-Science}%
  \BibitemOpen
  \bibfield  {author} {\bibinfo {author} {\bibfnamefont {L.~F.}\ \bibnamefont
  {Gomez~$\emph{et al.}$}},\ }\bibfield  {title} {\bibinfo {title} {Shapes and
  vorticities of superfluid helium nanodroplets},\ }\href
  {https://doi.org/10.1126/science.1252395} {\bibfield  {journal} {\bibinfo
  {journal} {Science}\ }\textbf {\bibinfo {volume} {345}},\ \bibinfo {pages}
  {906} (\bibinfo {year} {2014})}\BibitemShut {NoStop}%
\bibitem [{\citenamefont {Bewley}\ \emph {et~al.}(2006)\citenamefont {Bewley},
  \citenamefont {Lathrop},\ and\ \citenamefont
  {Sreenivasan}}]{Bewley-2006-Nature}%
  \BibitemOpen
  \bibfield  {author} {\bibinfo {author} {\bibfnamefont {G.~P.}\ \bibnamefont
  {Bewley}}, \bibinfo {author} {\bibfnamefont {D.~P.}\ \bibnamefont
  {Lathrop}},\ and\ \bibinfo {author} {\bibfnamefont {K.~R.}\ \bibnamefont
  {Sreenivasan}},\ }\bibfield  {title} {\bibinfo {title} {Superfluid helium:
  Visualization of quantized vortices},\ }\href@noop {} {\bibfield  {journal}
  {\bibinfo  {journal} {Nature}\ }\textbf {\bibinfo {volume} {441}},\ \bibinfo
  {pages} {588} (\bibinfo {year} {2006})}\BibitemShut {NoStop}%
\bibitem [{\citenamefont {Zmeev}\ \emph {et~al.}(2013)\citenamefont {Zmeev},
  \citenamefont {Pakpour}, \citenamefont {Walmsley}, \citenamefont {Golov},
  \citenamefont {Guo}, \citenamefont {McKinsey}, \citenamefont {Ihas},
  \citenamefont {McClintock}, \citenamefont {Fisher},\ and\ \citenamefont
  {Vinen}}]{Zmeev-2013-PRL}%
  \BibitemOpen
  \bibfield  {author} {\bibinfo {author} {\bibfnamefont {D.~E.}\ \bibnamefont
  {Zmeev}}, \bibinfo {author} {\bibfnamefont {F.}~\bibnamefont {Pakpour}},
  \bibinfo {author} {\bibfnamefont {P.~M.}\ \bibnamefont {Walmsley}}, \bibinfo
  {author} {\bibfnamefont {A.~I.}\ \bibnamefont {Golov}}, \bibinfo {author}
  {\bibfnamefont {W.}~\bibnamefont {Guo}}, \bibinfo {author} {\bibfnamefont
  {D.~N.}\ \bibnamefont {McKinsey}}, \bibinfo {author} {\bibfnamefont {G.~G.}\
  \bibnamefont {Ihas}}, \bibinfo {author} {\bibfnamefont {P.~V.~E.}\
  \bibnamefont {McClintock}}, \bibinfo {author} {\bibfnamefont {S.~N.}\
  \bibnamefont {Fisher}},\ and\ \bibinfo {author} {\bibfnamefont {W.~F.}\
  \bibnamefont {Vinen}},\ }\bibfield  {title} {\bibinfo {title} {Excimers
  ${\mathrm{he}}_{2}^{*}$ as tracers of quantum turbulence in $^{4}\mathrm{He}$
  in the $t=0$ limit},\ }\href {https://doi.org/10.1103/PhysRevLett.110.175303}
  {\bibfield  {journal} {\bibinfo  {journal} {Phys. Rev. Lett.}\ }\textbf
  {\bibinfo {volume} {110}},\ \bibinfo {pages} {175303} (\bibinfo {year}
  {2013})}\BibitemShut {NoStop}%
\bibitem [{\citenamefont {Mastracci}\ and\ \citenamefont
  {Guo}(2019)}]{Mastracci-2019-PRF}%
  \BibitemOpen
  \bibfield  {author} {\bibinfo {author} {\bibfnamefont {B.}~\bibnamefont
  {Mastracci}}\ and\ \bibinfo {author} {\bibfnamefont {W.}~\bibnamefont
  {Guo}},\ }\bibfield  {title} {\bibinfo {title} {Characterizing vortex tangle
  properties in steady-state {He II} counterflow using particle tracking
  velocimetry},\ }\href@noop {} {\bibfield  {journal} {\bibinfo  {journal}
  {Phys. Rev. Fluids}\ }\textbf {\bibinfo {volume} {4}},\ \bibinfo {pages}
  {023301} (\bibinfo {year} {2019})}\BibitemShut {NoStop}%
\bibitem [{\citenamefont {Toennies}\ \emph {et~al.}(2001)\citenamefont
  {Toennies}, \citenamefont {Vilesov},\ and\ \citenamefont
  {Whaley}}]{Toennies-2001-PT}%
  \BibitemOpen
  \bibfield  {author} {\bibinfo {author} {\bibfnamefont {J.~P.}\ \bibnamefont
  {Toennies}}, \bibinfo {author} {\bibfnamefont {A.~F.}\ \bibnamefont
  {Vilesov}},\ and\ \bibinfo {author} {\bibfnamefont {K.~B.}\ \bibnamefont
  {Whaley}},\ }\bibfield  {title} {\bibinfo {title} {Superfluid helium
  droplets: An ultracold nanolaboratory},\ }\href
  {https://doi.org/10.1063/1.1359707} {\bibfield  {journal} {\bibinfo
  {journal} {Phys. Today}\ }\textbf {\bibinfo {volume} {54}},\ \bibinfo {pages}
  {31} (\bibinfo {year} {2001})}\BibitemShut {NoStop}%
\bibitem [{\citenamefont {Latimer}\ \emph {et~al.}(2014)\citenamefont
  {Latimer}, \citenamefont {Spence}, \citenamefont {Feng}, \citenamefont
  {Boatwright}, \citenamefont {Ellis},\ and\ \citenamefont
  {Yang}}]{Latimer-2014-NL}%
  \BibitemOpen
  \bibfield  {author} {\bibinfo {author} {\bibfnamefont {E.}~\bibnamefont
  {Latimer}}, \bibinfo {author} {\bibfnamefont {D.}~\bibnamefont {Spence}},
  \bibinfo {author} {\bibfnamefont {C.}~\bibnamefont {Feng}}, \bibinfo {author}
  {\bibfnamefont {A.}~\bibnamefont {Boatwright}}, \bibinfo {author}
  {\bibfnamefont {A.~M.}\ \bibnamefont {Ellis}},\ and\ \bibinfo {author}
  {\bibfnamefont {S.}~\bibnamefont {Yang}},\ }\bibfield  {title} {\bibinfo
  {title} {Preparation of ultrathin nanowires using superfluid helium
  droplets},\ }\href {https://doi.org/10.1021/nl500946u} {\bibfield  {journal}
  {\bibinfo  {journal} {Nano Lett.}\ }\textbf {\bibinfo {volume} {14}},\
  \bibinfo {pages} {2902} (\bibinfo {year} {2014})}\BibitemShut {NoStop}%
\bibitem [{\citenamefont {Silk}\ and\ \citenamefont
  {Vilenkin}(1984)}]{Silk-1984-PRL}%
  \BibitemOpen
  \bibfield  {author} {\bibinfo {author} {\bibfnamefont {J.}~\bibnamefont
  {Silk}}\ and\ \bibinfo {author} {\bibfnamefont {A.}~\bibnamefont
  {Vilenkin}},\ }\bibfield  {title} {\bibinfo {title} {Cosmic strings and
  galaxy formation},\ }\href {https://doi.org/10.1103/PhysRevLett.53.1700}
  {\bibfield  {journal} {\bibinfo  {journal} {Phys. Rev. Lett.}\ }\textbf
  {\bibinfo {volume} {53}},\ \bibinfo {pages} {1700} (\bibinfo {year}
  {1984})}\BibitemShut {NoStop}%
\bibitem [{\citenamefont {Tsubota}\ \emph {et~al.}(2003)\citenamefont
  {Tsubota}, \citenamefont {Araki},\ and\ \citenamefont
  {Vinen}}]{Tsubota-2003-PB}%
  \BibitemOpen
  \bibfield  {author} {\bibinfo {author} {\bibfnamefont {M.}~\bibnamefont
  {Tsubota}}, \bibinfo {author} {\bibfnamefont {T.}~\bibnamefont {Araki}},\
  and\ \bibinfo {author} {\bibfnamefont {W.}~\bibnamefont {Vinen}},\ }\bibfield
   {title} {\bibinfo {title} {Diffusion of an inhomogeneous vortex tangle},\
  }\href {https://doi.org/https://doi.org/10.1016/S0921-4526(02)01968-3}
  {\bibfield  {journal} {\bibinfo  {journal} {Phys. B: Condens. Matter}\
  }\textbf {\bibinfo {volume} {329-333}},\ \bibinfo {pages} {224} (\bibinfo
  {year} {2003})},\ \bibinfo {note} {proceedings of the 23rd International
  Conference on Low Temperature Physics}\BibitemShut {NoStop}%
\bibitem [{\citenamefont {Rickinson}\ \emph {et~al.}(2019)\citenamefont
  {Rickinson}, \citenamefont {Parker}, \citenamefont {Baggaley},\ and\
  \citenamefont {Barenghi}}]{Rickinson-2019-PRB}%
  \BibitemOpen
  \bibfield  {author} {\bibinfo {author} {\bibfnamefont {E.}~\bibnamefont
  {Rickinson}}, \bibinfo {author} {\bibfnamefont {N.~G.}\ \bibnamefont
  {Parker}}, \bibinfo {author} {\bibfnamefont {A.~W.}\ \bibnamefont
  {Baggaley}},\ and\ \bibinfo {author} {\bibfnamefont {C.~F.}\ \bibnamefont
  {Barenghi}},\ }\bibfield  {title} {\bibinfo {title} {Inviscid diffusion of
  vorticity in low-temperature superfluid helium},\ }\href
  {https://doi.org/10.1103/PhysRevB.99.224501} {\bibfield  {journal} {\bibinfo
  {journal} {Phys. Rev. B}\ }\textbf {\bibinfo {volume} {99}},\ \bibinfo
  {pages} {224501} (\bibinfo {year} {2019})}\BibitemShut {NoStop}%
\bibitem [{\citenamefont {Tang}\ \emph {et~al.}(2021)\citenamefont {Tang},
  \citenamefont {Bao},\ and\ \citenamefont {Guo}}]{Tang-2021-PNAS}%
  \BibitemOpen
  \bibfield  {author} {\bibinfo {author} {\bibfnamefont {Y.}~\bibnamefont
  {Tang}}, \bibinfo {author} {\bibfnamefont {S.}~\bibnamefont {Bao}},\ and\
  \bibinfo {author} {\bibfnamefont {W.}~\bibnamefont {Guo}},\ }\bibfield
  {title} {\bibinfo {title} {Superdiffusion of quantized vortices uncovering
  scaling laws in quantum turbulence},\ }\bibfield  {journal} {\bibinfo
  {journal} {Proc. Natl. Acad. Sci. U.S.A}\ }\textbf {\bibinfo {volume}
  {118}},\ \href {https://doi.org/10.1073/pnas.2021957118}
  {10.1073/pnas.2021957118} (\bibinfo {year} {2021})\BibitemShut {NoStop}%
\bibitem [{\citenamefont {Adachi}\ \emph {et~al.}(2010)\citenamefont {Adachi},
  \citenamefont {Fujiyama},\ and\ \citenamefont {Tsubota}}]{Adachi-2010-PRB}%
  \BibitemOpen
  \bibfield  {author} {\bibinfo {author} {\bibfnamefont {H.}~\bibnamefont
  {Adachi}}, \bibinfo {author} {\bibfnamefont {S.}~\bibnamefont {Fujiyama}},\
  and\ \bibinfo {author} {\bibfnamefont {M.}~\bibnamefont {Tsubota}},\
  }\bibfield  {title} {\bibinfo {title} {Steady-state counterflow quantum
  turbulence: Simulation of vortex filaments using the full biot-savart law},\
  }\href {https://doi.org/10.1103/PhysRevB.81.104511} {\bibfield  {journal}
  {\bibinfo  {journal} {Phys. Rev. B}\ }\textbf {\bibinfo {volume} {81}},\
  \bibinfo {pages} {104511} (\bibinfo {year} {2010})}\BibitemShut {NoStop}%
\bibitem [{\citenamefont {Volovik}(2003)}]{Volovik-2003-JETP}%
  \BibitemOpen
  \bibfield  {author} {\bibinfo {author} {\bibfnamefont {G.}~\bibnamefont
  {Volovik}},\ }\bibfield  {title} {\bibinfo {title} {Classical and quantum
  regimes of superfluid turbulence},\ }\href
  {https://doi.org/10.1134/1.1641478} {\bibfield  {journal} {\bibinfo
  {journal} {Jetp Lett.}\ }\textbf {\bibinfo {volume} {78}},\ \bibinfo {pages}
  {533} (\bibinfo {year} {2003})}\BibitemShut {NoStop}%
\bibitem [{\citenamefont {Walmsley}\ and\ \citenamefont
  {Golov}(2008)}]{Walmsley-2008-PRL}%
  \BibitemOpen
  \bibfield  {author} {\bibinfo {author} {\bibfnamefont {P.~M.}\ \bibnamefont
  {Walmsley}}\ and\ \bibinfo {author} {\bibfnamefont {A.~I.}\ \bibnamefont
  {Golov}},\ }\bibfield  {title} {\bibinfo {title} {Quantum and quasiclassical
  types of superfluid turbulence},\ }\href
  {https://doi.org/10.1103/PhysRevLett.100.245301} {\bibfield  {journal}
  {\bibinfo  {journal} {Phys. Rev. Lett.}\ }\textbf {\bibinfo {volume} {100}},\
  \bibinfo {pages} {245301} (\bibinfo {year} {2008})}\BibitemShut {NoStop}%
\bibitem [{\citenamefont {Vinen}(1957{\natexlab{a}})}]{Vinen-1957-PRS-I}%
  \BibitemOpen
  \bibfield  {author} {\bibinfo {author} {\bibfnamefont {W.~F.}\ \bibnamefont
  {Vinen}},\ }\bibfield  {title} {\bibinfo {title} {Mutual friction in a heat
  current in liquid helium {II}. {I}. experiments on steady heat currents},\
  }\href {https://doi.org/10.1098/rspa.1957.0071} {\bibfield  {journal}
  {\bibinfo  {journal} {Proc. Roy. Soc. A}\ }\textbf {\bibinfo {volume}
  {240}},\ \bibinfo {pages} {114} (\bibinfo {year}
  {1957}{\natexlab{a}})}\BibitemShut {NoStop}%
\bibitem [{\citenamefont {Gao}\ \emph {et~al.}(2016{\natexlab{a}})\citenamefont
  {Gao}, \citenamefont {Guo}, \citenamefont {L'vov}, \citenamefont {Pomyalov},
  \citenamefont {Skrbek}, \citenamefont {Varga},\ and\ \citenamefont
  {Vinen}}]{Gao-2016-JETP}%
  \BibitemOpen
  \bibfield  {author} {\bibinfo {author} {\bibfnamefont {J.}~\bibnamefont
  {Gao}}, \bibinfo {author} {\bibfnamefont {W.}~\bibnamefont {Guo}}, \bibinfo
  {author} {\bibfnamefont {V.~S.}\ \bibnamefont {L'vov}}, \bibinfo {author}
  {\bibfnamefont {A.}~\bibnamefont {Pomyalov}}, \bibinfo {author}
  {\bibfnamefont {L.}~\bibnamefont {Skrbek}}, \bibinfo {author} {\bibfnamefont
  {E.}~\bibnamefont {Varga}},\ and\ \bibinfo {author} {\bibfnamefont {W.~F.}\
  \bibnamefont {Vinen}},\ }\bibfield  {title} {\bibinfo {title} {Decay of
  counterflow turbulence in superfluid ${^4}${H}e},\ }\href@noop {} {\bibfield
  {journal} {\bibinfo  {journal} {JETP Lett.}\ }\textbf {\bibinfo {volume}
  {103}},\ \bibinfo {pages} {648} (\bibinfo {year}
  {2016}{\natexlab{a}})}\BibitemShut {NoStop}%
\bibitem [{\citenamefont {Takeuchi}\ \emph {et~al.}(2010)\citenamefont
  {Takeuchi}, \citenamefont {Ishino},\ and\ \citenamefont
  {Tsubota}}]{Takeuchi-2010-PRL}%
  \BibitemOpen
  \bibfield  {author} {\bibinfo {author} {\bibfnamefont {H.}~\bibnamefont
  {Takeuchi}}, \bibinfo {author} {\bibfnamefont {S.}~\bibnamefont {Ishino}},\
  and\ \bibinfo {author} {\bibfnamefont {M.}~\bibnamefont {Tsubota}},\
  }\bibfield  {title} {\bibinfo {title} {Binary quantum turbulence arising from
  countersuperflow instability in two-component bose-einstein condensates},\
  }\href {https://doi.org/10.1103/PhysRevLett.105.205301} {\bibfield  {journal}
  {\bibinfo  {journal} {Phys. Rev. Lett.}\ }\textbf {\bibinfo {volume} {105}},\
  \bibinfo {pages} {205301} (\bibinfo {year} {2010})}\BibitemShut {NoStop}%
\bibitem [{\citenamefont {Greenstein}(1970)}]{Greenstein-1970-Nature}%
  \BibitemOpen
  \bibfield  {author} {\bibinfo {author} {\bibfnamefont {G.}~\bibnamefont
  {Greenstein}},\ }\bibfield  {title} {\bibinfo {title} {Superfluid turbulence
  in neutron stars},\ }\href@noop {} {\bibfield  {journal} {\bibinfo  {journal}
  {Nature}\ }\textbf {\bibinfo {volume} {227}},\ \bibinfo {pages} {791}
  (\bibinfo {year} {1970})}\BibitemShut {NoStop}%
\bibitem [{\citenamefont {Haskell}\ \emph {et~al.}(2020)\citenamefont
  {Haskell}, \citenamefont {Antonopoulou},\ and\ \citenamefont
  {Barenghi}}]{Haskell-2020-MNRAS}%
  \BibitemOpen
  \bibfield  {author} {\bibinfo {author} {\bibfnamefont {B.}~\bibnamefont
  {Haskell}}, \bibinfo {author} {\bibfnamefont {D.}~\bibnamefont
  {Antonopoulou}},\ and\ \bibinfo {author} {\bibfnamefont {C.}~\bibnamefont
  {Barenghi}},\ }\bibfield  {title} {\bibinfo {title} {{Turbulent, pinned
  superfluids in neutron stars and pulsar glitch recoveries}},\ }\href
  {https://doi.org/10.1093/mnras/staa2678} {\bibfield  {journal} {\bibinfo
  {journal} {Mon. Notices Royal Astron. Soc.}\ }\textbf {\bibinfo {volume}
  {499}},\ \bibinfo {pages} {161} (\bibinfo {year} {2020})}\BibitemShut
  {NoStop}%
\bibitem [{\citenamefont {Hartman}\ \emph {et~al.}(2021)\citenamefont
  {Hartman}, \citenamefont {Winther},\ and\ \citenamefont
  {Mota}}]{Hartman-2021-AA}%
  \BibitemOpen
  \bibfield  {author} {\bibinfo {author} {\bibfnamefont {S.~T.~H.}\
  \bibnamefont {Hartman}}, \bibinfo {author} {\bibfnamefont {H.~A.}\
  \bibnamefont {Winther}},\ and\ \bibinfo {author} {\bibfnamefont {D.~F.}\
  \bibnamefont {Mota}},\ }\bibfield  {title} {\bibinfo {title} {Dynamical
  friction in bose-einstein condensed self-interacting dark matter at finite
  temperatures, and the fornax dwarf spheroidal},\ }\href
  {https://doi.org/10.1051/0004-6361/202039865} {\bibfield  {journal} {\bibinfo
   {journal} {Astron. Astrophys.}\ }\textbf {\bibinfo {volume} {647}},\
  \bibinfo {pages} {A70} (\bibinfo {year} {2021})}\BibitemShut {NoStop}%
\bibitem [{\citenamefont {Sikivie}\ and\ \citenamefont
  {Yang}(2009)}]{Sikivie-2009-PRL}%
  \BibitemOpen
  \bibfield  {author} {\bibinfo {author} {\bibfnamefont {P.}~\bibnamefont
  {Sikivie}}\ and\ \bibinfo {author} {\bibfnamefont {Q.}~\bibnamefont {Yang}},\
  }\bibfield  {title} {\bibinfo {title} {Bose-einstein condensation of dark
  matter axions},\ }\href {https://doi.org/10.1103/PhysRevLett.103.111301}
  {\bibfield  {journal} {\bibinfo  {journal} {Phys. Rev. Lett.}\ }\textbf
  {\bibinfo {volume} {103}},\ \bibinfo {pages} {111301} (\bibinfo {year}
  {2009})}\BibitemShut {NoStop}%
\bibitem [{\citenamefont {Zurek}(1985)}]{Zurek-1985-Nature}%
  \BibitemOpen
  \bibfield  {author} {\bibinfo {author} {\bibfnamefont {W.~H.}\ \bibnamefont
  {Zurek}},\ }\bibfield  {title} {\bibinfo {title} {Cosmological experiments in
  superfluid-helium},\ }\href {https://doi.org/10.1038/317505a0} {\bibfield
  {journal} {\bibinfo  {journal} {Nature}\ }\textbf {\bibinfo {volume} {317}},\
  \bibinfo {pages} {505} (\bibinfo {year} {1985})}\BibitemShut {NoStop}%
\bibitem [{\citenamefont {Stagg}\ \emph {et~al.}(2016)\citenamefont {Stagg},
  \citenamefont {Parker},\ and\ \citenamefont {Barenghi}}]{Stagg-2016-PRA}%
  \BibitemOpen
  \bibfield  {author} {\bibinfo {author} {\bibfnamefont {G.~W.}\ \bibnamefont
  {Stagg}}, \bibinfo {author} {\bibfnamefont {N.~G.}\ \bibnamefont {Parker}},\
  and\ \bibinfo {author} {\bibfnamefont {C.~F.}\ \bibnamefont {Barenghi}},\
  }\bibfield  {title} {\bibinfo {title} {Ultraquantum turbulence in a quenched
  homogeneous bose gas},\ }\href {https://doi.org/10.1103/PhysRevA.94.053632}
  {\bibfield  {journal} {\bibinfo  {journal} {Phys. Rev. A}\ }\textbf {\bibinfo
  {volume} {94}},\ \bibinfo {pages} {053632} (\bibinfo {year}
  {2016})}\BibitemShut {NoStop}%
\bibitem [{\citenamefont {Landau}\ and\ \citenamefont
  {Lifshitz}(1987)}]{Landau-book}%
  \BibitemOpen
  \bibfield  {author} {\bibinfo {author} {\bibfnamefont {L.~D.}\ \bibnamefont
  {Landau}}\ and\ \bibinfo {author} {\bibfnamefont {E.~M.}\ \bibnamefont
  {Lifshitz}},\ }\href {https://doi.org/10.1016/C2013-0-03799-1} {\emph
  {\bibinfo {title} {Fluid Mechanics}}},\ \bibinfo {edition} {2nd}\ ed.,\
  Vol.~\bibinfo {volume} {6}\ (\bibinfo  {publisher} {{Pergamon Press}},\
  \bibinfo {address} {Oxford},\ \bibinfo {year} {1987})\BibitemShut {NoStop}%
\bibitem [{\citenamefont {Schwarz}(1988)}]{Schwarz-1988-PRB}%
  \BibitemOpen
  \bibfield  {author} {\bibinfo {author} {\bibfnamefont {K.~W.}\ \bibnamefont
  {Schwarz}},\ }\bibfield  {title} {\bibinfo {title} {Three-dimensional vortex
  dynamics in superfluid $^{4}\mathrm{He}$: Homogeneous superfluid
  turbulence},\ }\href {https://doi.org/10.1103/PhysRevB.38.2398} {\bibfield
  {journal} {\bibinfo  {journal} {Phys. Rev. B}\ }\textbf {\bibinfo {volume}
  {38}},\ \bibinfo {pages} {2398} (\bibinfo {year} {1988})}\BibitemShut
  {NoStop}%
\bibitem [{\citenamefont {Tsubota}\ \emph {et~al.}(2000)\citenamefont
  {Tsubota}, \citenamefont {Araki},\ and\ \citenamefont
  {Nemirovskii}}]{Tsubota-2000-PRB}%
  \BibitemOpen
  \bibfield  {author} {\bibinfo {author} {\bibfnamefont {M.}~\bibnamefont
  {Tsubota}}, \bibinfo {author} {\bibfnamefont {T.}~\bibnamefont {Araki}},\
  and\ \bibinfo {author} {\bibfnamefont {S.~K.}\ \bibnamefont {Nemirovskii}},\
  }\bibfield  {title} {\bibinfo {title} {Dynamics of vortex tangle without
  mutual friction in superfluid ${}^{4}\mathrm{He}$},\ }\href
  {https://doi.org/10.1103/PhysRevB.62.11751} {\bibfield  {journal} {\bibinfo
  {journal} {Phys. Rev. B}\ }\textbf {\bibinfo {volume} {62}},\ \bibinfo
  {pages} {11751} (\bibinfo {year} {2000})}\BibitemShut {NoStop}%
\bibitem [{\citenamefont {Baggaley}(2012)}]{Baggaley-2012-JLTP}%
  \BibitemOpen
  \bibfield  {author} {\bibinfo {author} {\bibfnamefont {A.~W.}\ \bibnamefont
  {Baggaley}},\ }\bibfield  {title} {\bibinfo {title} {The sensitivity of the
  vortex filament method to different reconnection models},\ }\href
  {https://doi.org/10.1007/s10909-012-0605-8} {\bibfield  {journal} {\bibinfo
  {journal} {J. Low Temp. Phys.}\ }\textbf {\bibinfo {volume} {168}},\ \bibinfo
  {pages} {18} (\bibinfo {year} {2012})}\BibitemShut {NoStop}%
\bibitem [{\citenamefont {Ben-Avraham}\ and\ \citenamefont
  {Havlin}(2000)}]{Ben-2000-book}%
  \BibitemOpen
  \bibfield  {author} {\bibinfo {author} {\bibfnamefont {D.}~\bibnamefont
  {Ben-Avraham}}\ and\ \bibinfo {author} {\bibfnamefont {S.}~\bibnamefont
  {Havlin}},\ }\href@noop {} {\emph {\bibinfo {title} {Diffusion and Reactions
  in Fractals and Disordered Systems}}}\ (\bibinfo  {publisher} {Cambridge
  University Press},\ \bibinfo {address} {Cambridge, United Kingdom},\ \bibinfo
  {year} {2000})\BibitemShut {NoStop}%
\bibitem [{\citenamefont {Solomon}\ \emph {et~al.}(1993)\citenamefont
  {Solomon}, \citenamefont {Weeks},\ and\ \citenamefont
  {Swinney}}]{Solomon-1993-PRL}%
  \BibitemOpen
  \bibfield  {author} {\bibinfo {author} {\bibfnamefont {T.~H.}\ \bibnamefont
  {Solomon}}, \bibinfo {author} {\bibfnamefont {E.~R.}\ \bibnamefont {Weeks}},\
  and\ \bibinfo {author} {\bibfnamefont {H.~L.}\ \bibnamefont {Swinney}},\
  }\bibfield  {title} {\bibinfo {title} {Observation of anomalous diffusion and
  l{\'e}vy flights in a two-dimensional rotating flow},\ }\href@noop {}
  {\bibfield  {journal} {\bibinfo  {journal} {Phys. Rev. Lett.}\ }\textbf
  {\bibinfo {volume} {71}},\ \bibinfo {pages} {3975} (\bibinfo {year}
  {1993})}\BibitemShut {NoStop}%
\bibitem [{\citenamefont {Harris}\ \emph {et~al.}(2012)\citenamefont {Harris},
  \citenamefont {Banigan}, \citenamefont {Christian}, \citenamefont {Konradt},
  \citenamefont {Wojno}, \citenamefont {Norose}, \citenamefont {Wilson},
  \citenamefont {John}, \citenamefont {Weninger}, \citenamefont {Luster} \emph
  {et~al.}}]{Harris-2012-Nature}%
  \BibitemOpen
  \bibfield  {author} {\bibinfo {author} {\bibfnamefont {T.~H.}\ \bibnamefont
  {Harris}}, \bibinfo {author} {\bibfnamefont {E.~J.}\ \bibnamefont {Banigan}},
  \bibinfo {author} {\bibfnamefont {D.~A.}\ \bibnamefont {Christian}}, \bibinfo
  {author} {\bibfnamefont {C.}~\bibnamefont {Konradt}}, \bibinfo {author}
  {\bibfnamefont {E.~D.~T.}\ \bibnamefont {Wojno}}, \bibinfo {author}
  {\bibfnamefont {K.}~\bibnamefont {Norose}}, \bibinfo {author} {\bibfnamefont
  {E.~H.}\ \bibnamefont {Wilson}}, \bibinfo {author} {\bibfnamefont
  {B.}~\bibnamefont {John}}, \bibinfo {author} {\bibfnamefont {W.}~\bibnamefont
  {Weninger}}, \bibinfo {author} {\bibfnamefont {A.~D.}\ \bibnamefont
  {Luster}}, \emph {et~al.},\ }\bibfield  {title} {\bibinfo {title}
  {Generalized l{\'e}vy walks and the role of chemokines in migration of
  effector {CD8+ T} cells},\ }\href@noop {} {\bibfield  {journal} {\bibinfo
  {journal} {Nature}\ }\textbf {\bibinfo {volume} {486}},\ \bibinfo {pages}
  {545} (\bibinfo {year} {2012})}\BibitemShut {NoStop}%
\bibitem [{\citenamefont {Sagi}\ \emph {et~al.}(2012)\citenamefont {Sagi},
  \citenamefont {Brook}, \citenamefont {Almog},\ and\ \citenamefont
  {Davidson}}]{Sagi-2012-PRL}%
  \BibitemOpen
  \bibfield  {author} {\bibinfo {author} {\bibfnamefont {Y.}~\bibnamefont
  {Sagi}}, \bibinfo {author} {\bibfnamefont {M.}~\bibnamefont {Brook}},
  \bibinfo {author} {\bibfnamefont {I.}~\bibnamefont {Almog}},\ and\ \bibinfo
  {author} {\bibfnamefont {N.}~\bibnamefont {Davidson}},\ }\bibfield  {title}
  {\bibinfo {title} {Observation of anomalous diffusion and fractional
  self-similarity in one dimension},\ }\href@noop {} {\bibfield  {journal}
  {\bibinfo  {journal} {Phys. Rev. Lett.}\ }\textbf {\bibinfo {volume} {108}},\
  \bibinfo {pages} {093002} (\bibinfo {year} {2012})}\BibitemShut {NoStop}%
\bibitem [{\citenamefont {Raichlen}\ \emph {et~al.}(2014)\citenamefont
  {Raichlen}, \citenamefont {Wood}, \citenamefont {Gordon}, \citenamefont
  {Mabulla}, \citenamefont {Marlowe},\ and\ \citenamefont
  {Pontzer}}]{Raichlen-2014-PNAS}%
  \BibitemOpen
  \bibfield  {author} {\bibinfo {author} {\bibfnamefont {D.~A.}\ \bibnamefont
  {Raichlen}}, \bibinfo {author} {\bibfnamefont {B.~M.}\ \bibnamefont {Wood}},
  \bibinfo {author} {\bibfnamefont {A.~D.}\ \bibnamefont {Gordon}}, \bibinfo
  {author} {\bibfnamefont {A.~Z.}\ \bibnamefont {Mabulla}}, \bibinfo {author}
  {\bibfnamefont {F.~W.}\ \bibnamefont {Marlowe}},\ and\ \bibinfo {author}
  {\bibfnamefont {H.}~\bibnamefont {Pontzer}},\ }\bibfield  {title} {\bibinfo
  {title} {Evidence of l{\'e}vy walk foraging patterns in human
  hunter--gatherers},\ }\href@noop {} {\bibfield  {journal} {\bibinfo
  {journal} {Proc. Natl. Acad. Sci.}\ }\textbf {\bibinfo {volume} {111}},\
  \bibinfo {pages} {728} (\bibinfo {year} {2014})}\BibitemShut {NoStop}%
\bibitem [{\citenamefont {Bouchaud}\ and\ \citenamefont
  {Georges}(1990)}]{Bouchaud-1990-PR}%
  \BibitemOpen
  \bibfield  {author} {\bibinfo {author} {\bibfnamefont {J.-P.}\ \bibnamefont
  {Bouchaud}}\ and\ \bibinfo {author} {\bibfnamefont {A.}~\bibnamefont
  {Georges}},\ }\bibfield  {title} {\bibinfo {title} {Anomalous diffusion in
  disordered media: Statistical mechanisms, models and physical applications},\
  }\href {https://doi.org/https://doi.org/10.1016/0370-1573(90)90099-N}
  {\bibfield  {journal} {\bibinfo  {journal} {Phys. Rep.}\ }\textbf {\bibinfo
  {volume} {195}},\ \bibinfo {pages} {127} (\bibinfo {year}
  {1990})}\BibitemShut {NoStop}%
\bibitem [{\citenamefont {Bewley}\ \emph {et~al.}(2008)\citenamefont {Bewley},
  \citenamefont {Paoletti}, \citenamefont {Sreenivasan},\ and\ \citenamefont
  {Lathrop}}]{Bewley-2008-PNAS}%
  \BibitemOpen
  \bibfield  {author} {\bibinfo {author} {\bibfnamefont {G.~P.}\ \bibnamefont
  {Bewley}}, \bibinfo {author} {\bibfnamefont {M.~S.}\ \bibnamefont
  {Paoletti}}, \bibinfo {author} {\bibfnamefont {K.~R.}\ \bibnamefont
  {Sreenivasan}},\ and\ \bibinfo {author} {\bibfnamefont {D.~P.}\ \bibnamefont
  {Lathrop}},\ }\bibfield  {title} {\bibinfo {title} {Characterization of
  reconnecting vortices in superfluid helium},\ }\href
  {https://doi.org/10.1073/pnas.0806002105} {\bibfield  {journal} {\bibinfo
  {journal} {Proc. Natl. Acad. Sci. U.S.A}\ }\textbf {\bibinfo {volume}
  {105}},\ \bibinfo {pages} {13707} (\bibinfo {year} {2008})}\BibitemShut
  {NoStop}%
\bibitem [{\citenamefont {Paoletti}\ \emph {et~al.}(2008)\citenamefont
  {Paoletti}, \citenamefont {Fisher}, \citenamefont {Sreenivasan},\ and\
  \citenamefont {Lathrop}}]{Paoletti-2008-PRL}%
  \BibitemOpen
  \bibfield  {author} {\bibinfo {author} {\bibfnamefont {M.~S.}\ \bibnamefont
  {Paoletti}}, \bibinfo {author} {\bibfnamefont {M.~E.}\ \bibnamefont
  {Fisher}}, \bibinfo {author} {\bibfnamefont {K.~R.}\ \bibnamefont
  {Sreenivasan}},\ and\ \bibinfo {author} {\bibfnamefont {D.~P.}\ \bibnamefont
  {Lathrop}},\ }\bibfield  {title} {\bibinfo {title} {Velocity statistics
  distinguish quantum turbulence from classical turbulence},\ }\href@noop {}
  {\bibfield  {journal} {\bibinfo  {journal} {Phys. Rev. Lett.}\ }\textbf
  {\bibinfo {volume} {101}},\ \bibinfo {pages} {154501} (\bibinfo {year}
  {2008})}\BibitemShut {NoStop}%
\bibitem [{\citenamefont {Davison}\ and\ \citenamefont
  {Cox}(1989)}]{Davison-1989-PRS}%
  \BibitemOpen
  \bibfield  {author} {\bibinfo {author} {\bibfnamefont {A.~C.}\ \bibnamefont
  {Davison}}\ and\ \bibinfo {author} {\bibfnamefont {D.~R.}\ \bibnamefont
  {Cox}},\ }\bibfield  {title} {\bibinfo {title} {Some simple properties of
  sums of randoms variable having long-range dependence},\ }\href@noop {}
  {\bibfield  {journal} {\bibinfo  {journal} {Proc. R. Soc. Lond. A}\ }\textbf
  {\bibinfo {volume} {424}},\ \bibinfo {pages} {255} (\bibinfo {year}
  {1989})}\BibitemShut {NoStop}%
\bibitem [{\citenamefont {Mazzitelli}\ and\ \citenamefont
  {Lohse}(2004)}]{Mazzitelli-2004-NJP}%
  \BibitemOpen
  \bibfield  {author} {\bibinfo {author} {\bibfnamefont {I.~M.}\ \bibnamefont
  {Mazzitelli}}\ and\ \bibinfo {author} {\bibfnamefont {D.}~\bibnamefont
  {Lohse}},\ }\bibfield  {title} {\bibinfo {title} {Lagrangian statistics for
  fluid particles and bubbles in turbulence},\ }\href
  {https://doi.org/10.1088/1367-2630/6/1/203} {\bibfield  {journal} {\bibinfo
  {journal} {New J. Phys}\ }\textbf {\bibinfo {volume} {6}},\ \bibinfo {pages}
  {203} (\bibinfo {year} {2004})}\BibitemShut {NoStop}%
\bibitem [{\citenamefont {Vinen}(1957{\natexlab{b}})}]{Vinen-1957-PRS-III}%
  \BibitemOpen
  \bibfield  {author} {\bibinfo {author} {\bibfnamefont {W.~F.}\ \bibnamefont
  {Vinen}},\ }\bibfield  {title} {\bibinfo {title} {Mutual friction in a heat
  current in liquid helium {II}. {III}. theory of the mutual friction},\ }\href
  {https://doi.org/10.1098/rspa.1957.0191} {\bibfield  {journal} {\bibinfo
  {journal} {Proc. Roy. Soc. A}\ }\textbf {\bibinfo {volume} {242}},\ \bibinfo
  {pages} {493} (\bibinfo {year} {1957}{\natexlab{b}})}\BibitemShut {NoStop}%
\bibitem [{\citenamefont {Donnelly}(1991)}]{Donnelly-1991-B}%
  \BibitemOpen
  \bibfield  {author} {\bibinfo {author} {\bibfnamefont {R.~J.}\ \bibnamefont
  {Donnelly}},\ }\href@noop {} {\emph {\bibinfo {title} {Quantized vortices in
  helium II}}},\ Vol.~\bibinfo {volume} {2}\ (\bibinfo  {publisher} {Cambridge
  University Press},\ \bibinfo {year} {1991})\BibitemShut {NoStop}%
\bibitem [{\citenamefont {Donnelly}\ and\ \citenamefont
  {Barenghi}(1998)}]{Donnelly-1998-JPCRD}%
  \BibitemOpen
  \bibfield  {author} {\bibinfo {author} {\bibfnamefont {R.~J.}\ \bibnamefont
  {Donnelly}}\ and\ \bibinfo {author} {\bibfnamefont {C.~F.}\ \bibnamefont
  {Barenghi}},\ }\bibfield  {title} {\bibinfo {title} {The observed properties
  of liquid helium at the saturated vapor pressure},\ }\href
  {https://doi.org/10.1063/1.556028} {\bibfield  {journal} {\bibinfo  {journal}
  {J. Phys. Chem. Ref. Data}\ }\textbf {\bibinfo {volume} {27}},\ \bibinfo
  {pages} {1217} (\bibinfo {year} {1998})}\BibitemShut {NoStop}%
\bibitem [{\citenamefont {Kivotides}\ \emph {et~al.}(2001)\citenamefont
  {Kivotides}, \citenamefont {Vassilicos}, \citenamefont {Samuels},\ and\
  \citenamefont {Barenghi}}]{Kivotides-2001-PRL}%
  \BibitemOpen
  \bibfield  {author} {\bibinfo {author} {\bibfnamefont {D.}~\bibnamefont
  {Kivotides}}, \bibinfo {author} {\bibfnamefont {J.~C.}\ \bibnamefont
  {Vassilicos}}, \bibinfo {author} {\bibfnamefont {D.~C.}\ \bibnamefont
  {Samuels}},\ and\ \bibinfo {author} {\bibfnamefont {C.~F.}\ \bibnamefont
  {Barenghi}},\ }\bibfield  {title} {\bibinfo {title} {Kelvin waves cascade in
  superfluid turbulence},\ }\href {https://doi.org/10.1103/PhysRevLett.86.3080}
  {\bibfield  {journal} {\bibinfo  {journal} {Phys. Rev. Lett.}\ }\textbf
  {\bibinfo {volume} {86}},\ \bibinfo {pages} {3080} (\bibinfo {year}
  {2001})}\BibitemShut {NoStop}%
\bibitem [{\citenamefont {Gao}\ \emph {et~al.}(2016{\natexlab{b}})\citenamefont
  {Gao}, \citenamefont {Guo},\ and\ \citenamefont {Vinen}}]{Gao-2016-PRB}%
  \BibitemOpen
  \bibfield  {author} {\bibinfo {author} {\bibfnamefont {J.}~\bibnamefont
  {Gao}}, \bibinfo {author} {\bibfnamefont {W.}~\bibnamefont {Guo}},\ and\
  \bibinfo {author} {\bibfnamefont {W.~F.}\ \bibnamefont {Vinen}},\ }\bibfield
  {title} {\bibinfo {title} {Determination of the effective kinematic viscosity
  for the decay of quasiclassical turbulence in superfluid ${^4}${H}e},\
  }\href@noop {} {\bibfield  {journal} {\bibinfo  {journal} {Phys. Rev. B}\
  }\textbf {\bibinfo {volume} {94}},\ \bibinfo {pages} {094502} (\bibinfo
  {year} {2016}{\natexlab{b}})}\BibitemShut {NoStop}%
\bibitem [{\citenamefont {Gao}\ \emph {et~al.}(2018)\citenamefont {Gao},
  \citenamefont {Guo}, \citenamefont {Yui}, \citenamefont {Tsubota},\ and\
  \citenamefont {Vinen}}]{Gao-2018-PRB}%
  \BibitemOpen
  \bibfield  {author} {\bibinfo {author} {\bibfnamefont {J.}~\bibnamefont
  {Gao}}, \bibinfo {author} {\bibfnamefont {W.}~\bibnamefont {Guo}}, \bibinfo
  {author} {\bibfnamefont {S.}~\bibnamefont {Yui}}, \bibinfo {author}
  {\bibfnamefont {M.}~\bibnamefont {Tsubota}},\ and\ \bibinfo {author}
  {\bibfnamefont {W.~F.}\ \bibnamefont {Vinen}},\ }\bibfield  {title} {\bibinfo
  {title} {Dissipation in quantum turbulence in superfluid $^{4}\mathrm{He}$
  above 1 k},\ }\href {https://doi.org/10.1103/PhysRevB.97.184518} {\bibfield
  {journal} {\bibinfo  {journal} {Phys. Rev. B}\ }\textbf {\bibinfo {volume}
  {97}},\ \bibinfo {pages} {184518} (\bibinfo {year} {2018})}\BibitemShut
  {NoStop}%
\end{thebibliography}%


\begin{thebibliography}{10}
\expandafter\ifx\csname url\endcsname\relax
  \def\url#1{\texttt{#1}}\fi
\expandafter\ifx\csname urlprefix\endcsname\relax\def\urlprefix{URL }\fi
\providecommand{\bibinfo}[2]{#2}
\providecommand{\eprint}[2][]{\url{#2}}

\bibitem{Schwarz-1988-PRB}
\bibinfo{author}{Schwarz, K.~W.}
\newblock \bibinfo{title}{Three-dimensional vortex dynamics in superfluid
  $^{4}\mathrm{He}$: Homogeneous superfluid turbulence}.
\newblock \emph{\bibinfo{journal}{Phys. Rev. B}} \textbf{\bibinfo{volume}{38}},
  \bibinfo{pages}{2398--2417} (\bibinfo{year}{1988}).

\bibitem{Adachi-2010-PRB}
\bibinfo{author}{Adachi, H.}, \bibinfo{author}{Fujiyama, S.} \&
  \bibinfo{author}{Tsubota, M.}
\newblock \bibinfo{title}{Steady-state counterflow quantum turbulence:
  Simulation of vortex filaments using the full biot-savart law}.
\newblock \emph{\bibinfo{journal}{Phys. Rev. B}} \textbf{\bibinfo{volume}{81}},
  \bibinfo{pages}{104511} (\bibinfo{year}{2010}).

\bibitem{Press-1992-book}
\bibinfo{author}{Press, W.~H.}, \bibinfo{author}{Flannery, B.~P.},
  \bibinfo{author}{Teukolsky, S.~A.} \& \bibinfo{author}{Vetterling, W.~T.}
\newblock \emph{\bibinfo{title}{Numerical Recipes in C. The Art of Scientific
  Computing}} (\bibinfo{publisher}{{Cambridge University Press}},
  \bibinfo{address}{{Cambridge}}, \bibinfo{year}{1992}).

\bibitem{Tsubota-2000-PRB}
\bibinfo{author}{Tsubota, M.}, \bibinfo{author}{Araki, T.} \&
  \bibinfo{author}{Nemirovskii, S.~K.}
\newblock \bibinfo{title}{Dynamics of vortex tangle without mutual friction in
  superfluid ${}^{4}\mathrm{He}$}.
\newblock \emph{\bibinfo{journal}{Phys. Rev. B}} \textbf{\bibinfo{volume}{62}},
  \bibinfo{pages}{11751--11762} (\bibinfo{year}{2000}).

\bibitem{Baggaley-2012-JLTP}
\bibinfo{author}{Baggaley, A.~W.}
\newblock \bibinfo{title}{The sensitivity of the vortex filament method to
  different reconnection models}.
\newblock \emph{\bibinfo{journal}{J. Low Temp. Phys.}}
  \textbf{\bibinfo{volume}{168}}, \bibinfo{pages}{18--30}
  (\bibinfo{year}{2012}).

\bibitem{Zaburdaev-2015-RMP}
\bibinfo{author}{Zaburdaev, V.}, \bibinfo{author}{Denisov, S.} \&
  \bibinfo{author}{Klafter, J.}
\newblock \bibinfo{title}{L\'evy walks}.
\newblock \emph{\bibinfo{journal}{Rev. Mod. Phys.}}
  \textbf{\bibinfo{volume}{87}}, \bibinfo{pages}{483--530}
  (\bibinfo{year}{2015}).

\bibitem{Bouchaud-1990-PR}
\bibinfo{author}{Bouchaud, J.-P.} \& \bibinfo{author}{Georges, A.}
\newblock \bibinfo{title}{Anomalous diffusion in disordered media: Statistical
  mechanisms, models and physical applications}.
\newblock \emph{\bibinfo{journal}{Phys. Rep.}} \textbf{\bibinfo{volume}{195}},
  \bibinfo{pages}{127--293} (\bibinfo{year}{1990}).

\bibitem{Donnelly-1991-B}
\bibinfo{author}{Donnelly, R.~J.}
\newblock \emph{\bibinfo{title}{Quantized vortices in helium {II}}},
  vol.~\bibinfo{volume}{2} (\bibinfo{publisher}{Cambridge University Press},
  \bibinfo{year}{1991}).

\bibitem{Paoletti-2008-PRL}
\bibinfo{author}{Paoletti, M.~S.}, \bibinfo{author}{Fisher, M.~E.},
  \bibinfo{author}{Sreenivasan, K.~R.} \& \bibinfo{author}{Lathrop, D.~P.}
\newblock \bibinfo{title}{Velocity statistics distinguish quantum turbulence
  from classical turbulence}.
\newblock \emph{\bibinfo{journal}{Phys. Rev. Lett.}}
  \textbf{\bibinfo{volume}{101}}, \bibinfo{pages}{154501}
  (\bibinfo{year}{2008}).

\bibitem{Mastracci-2019-PRF}
\bibinfo{author}{Mastracci, B.} \& \bibinfo{author}{Guo, W.}
\newblock \bibinfo{title}{Characterizing vortex tangle properties in
  steady-state {He II} counterflow using particle tracking velocimetry}.
\newblock \emph{\bibinfo{journal}{Phys. Rev. Fluids}}
  \textbf{\bibinfo{volume}{4}}, \bibinfo{pages}{023301} (\bibinfo{year}{2019}).

\bibitem{White-2010-PRL}
\bibinfo{author}{White, A.~C.}, \bibinfo{author}{Barenghi, C.~F.},
  \bibinfo{author}{Proukakis, N.~P.}, \bibinfo{author}{Youd, A.~J.} \&
  \bibinfo{author}{Wacks, D.~H.}
\newblock \bibinfo{title}{Nonclassical velocity statistics in a turbulent
  atomic bose-einstein condensate}.
\newblock \emph{\bibinfo{journal}{Phys. Rev. Lett.}}
  \textbf{\bibinfo{volume}{104}}, \bibinfo{pages}{075301}
  (\bibinfo{year}{2010}).

\bibitem{Adachi-2011-PRB}
\bibinfo{author}{Adachi, H.} \& \bibinfo{author}{Tsubota, M.}
\newblock \bibinfo{title}{Numerical study of velocity statistics in steady
  counterflow quantum turbulence}.
\newblock \emph{\bibinfo{journal}{Phys. Rev. B}} \textbf{\bibinfo{volume}{83}},
  \bibinfo{pages}{132503} (\bibinfo{year}{2011}).

\bibitem{Tang-2021-PNAS}
\bibinfo{author}{Tang, Y.}, \bibinfo{author}{Bao, S.} \& \bibinfo{author}{Guo,
  W.}
\newblock \bibinfo{title}{Superdiffusion of quantized vortices uncovering
  scaling laws in quantum turbulence}.
\newblock \emph{\bibinfo{journal}{Proc. Natl. Acad. Sci. U.S.A}}
  \textbf{\bibinfo{volume}{118}} (\bibinfo{year}{2021}).

\end{thebibliography}

\end{document}


\title{Supplemental Materials for: Universal anomalous turbulent diffusion in quantum fluids}
\author{Satoshi Yui}
\thanks{Equal contributions}
\affiliation{Research and Education Center for Natural Sciences, Keio University, 4-1-1 Hiyoshi, Kohoku-ku, Yokohama 223-8521, Japan}

\author{Yuan Tang}
\thanks{Equal contributions}
\affiliation{National High Magnetic Field Laboratory, 1800 East Paul Dirac Drive, Tallahassee, Florida 32310, USA}
\affiliation{Mechanical Engineering Department, FAMU-FSU College of Engineering, Florida State University, Tallahassee, Florida 32310, USA}

\author{Wei Guo}
\email[Email: ]{wguo@magnet.fsu.edu}
\affiliation{National High Magnetic Field Laboratory, 1800 East Paul Dirac Drive, Tallahassee, Florida 32310, USA}
\affiliation{Mechanical Engineering Department, FAMU-FSU College of Engineering, Florida State University, Tallahassee, Florida 32310, USA}

\author{Hiromichi Kobayashi}
\email[Email: ]{hkobayas@keio.jp}
\affiliation{Research and Education Center for Natural Sciences, Keio University, 4-1-1 Hiyoshi, Kohoku-ku, Yokohama 223-8521, Japan}
\affiliation{Department of Physics, Keio University, 4-1-1 Hiyoshi, Kohoku-ku, Yokohama 223-8521, Japan}

\author{Makoto Tsubota}
\email[Email: ]{tsubota@osaka-cu.ac.jp}
\affiliation{Department of Physics \& Nambu Yoichiro Institute of Theoretical and Experimental Physics (NITEP) \& The OCU Advanced Research Institute for Natural Science and Technology (OCARINA), Osaka City University, 3-3-138 Sugimoto, Sumiyoshi-ku, Osaka 558-8585, Japan}

\maketitle

\section{Vortex filament method}
In the framework of the vortex filament model~\cite{Schwarz-1988-PRB}, all the quantized vortex lines are represented by zero-thickness filaments. These filaments are discretized into a series of points with a spatial separation $\Delta\xi$ in the range of $\Delta\xi_{min}=0.008$ mm to $\Delta\xi_{max}=0.024$ mm in our simulations. In the absence of the viscous normal fluid, a vortex-filament point at $\mathbf{s}$ moves at the local superfluid velocity $\mathbf{v}_s(\mathbf{s})$ as:
\begin{equation}
\frac{d\mathbf{s}}{dt}=\mathbf{v}_s(\mathbf{s})=\mathbf{v}_0(\mathbf{s})+\mathbf{v}_{in}(\mathbf{s}),
\end{equation}
where $\mathbf{v}_0(\mathbf{s})$ is the applied background superfluid velocity and $\mathbf{v}_{in}(\mathbf{s})$ denotes the velocity induced at $\mathbf{s}$ by all the vortices according to the Boit-Savart law~\cite{Schwarz-1988-PRB}:
\begin{equation}
\mathbf{v}_{in}(\mathbf{s})=\frac{\kappa}{4\pi}\int\frac{(\mathbf{s_1}-\mathbf{s})\times d\mathbf{s_1}}{|\mathbf{s_1}-\mathbf{s}|^3},
\end{equation}
where the integration is supposed to be performed along all the vortex filaments. However, since the integrant diverges at $\mathbf{s}$ for an ideal zero-thickness filament, we follow Adachi \emph{et al.}~\cite{Adachi-2010-PRB} and calculate the integral as the sum of the local contribution and the non-local contribution:
\begin{equation}
\mathbf{v}_{in}(\mathbf{s})=\beta_l\mathbf{s}'\times\mathbf{s}''+\frac{\kappa}{4\pi}\int'\frac{(\mathbf{s_1}-\mathbf{s})\times d\mathbf{s_1}}{|\mathbf{s_1}-\mathbf{s}|^3},
\end{equation}
where $\mathbf{s}'$ and $\mathbf{s}''$ in the local term are respectively the unit vectors along and perpendicular to the filament at $\mathbf{s}$, and the non-local term represents the integral along the rest of the filament and all other vortices. The coefficient $\beta_l$ is given by~\cite{Schwarz-1988-PRB}:
\begin{equation}
\beta_l=\frac{\kappa}{4\pi}\ln\left(\frac{2(l_+l_-)^{1/2}}{e^{1/4}a_0}\right)
\end{equation}
where $l_+$ and $l_-$ are the distances from the point at $\mathbf{s}$ to its two nearest neighbour points along the same filament, and the cut-off parameter $a_0\simeq1$~{\AA} denotes the vortex core radius in He II. At finite temperatures, the vortices also experience a mutual friction force from the viscous normal fluid. The equation of motion of the vortex-filament points is then given by Eq.~(3) in the main paper.

In our ring-injection simulations, the normal-fluid velocity is set to $\mathbf{v}_n=0$ and so is the background superfluid velocity $\mathbf{v}_0=0$. On the other hand, in the counterflow simulations we set $\mathbf{v}_n=U_n\mathbf{\hat{e}}_z$ and $\mathbf{v}_0=-U_s\mathbf{\hat{e}}_z$. The time evolution of the vortices can be obtained through a temporal integration of the Eq.~(1) or the Eq.~(3) in the main paper using the fourth-order Runge-Kutta method~\cite{Press-1992-book} with a time step $\Delta t=10^{-4}$ s. We have confirmed that our spatial and time resolutions are sufficient such that the simulation results are independent of the resolutions. In the evolution of the vortex tangle, whenever two vortex filaments approach to have a minimum separation less than $\Delta\xi_{min}$, we reconnect the two filaments at the location of the minimum separation following the procedures as detailed in ref.~\cite{Tsubota-2000-PRB,Baggaley-2012-JLTP}. Furthermore, to maintain the spatial resolution along the filaments, at each time step we delete (or insert) a vortex-filament point between any two adjacent points that have a separation $\Delta\xi<\Delta\xi_{min}$ (or $\Delta\xi>\Delta\xi_{max}$). When simulating the vortex tangle dynamics at 0 K, we also remove small vortex loops with lengths less than $5\Delta\xi_{min}$ to account for the cascade loss of the vortices~\cite{Tsubota-2000-PRB}.\\

\section{Vortex-point tracking}
In order to study the apparent diffusion of the vortex-filament points, we give every point a unique index so that it can be identified and tracked during its lifetime. However, tracking a vortex-filament point for long times (i.e., over a second) is very challenging. This is because as we adjust the spatial resolution along the vortex filaments at each time step, some vortex-filament points are removed and new points are added. A given vortex-filament point can hardly survive many time steps, which limits the number of samples at large diffusion times. To mitigate this issue, in our typical simulation we randomly tag 200 vortex-filament points in the tangle. When a tagged point and its nearest neighbour point have a separation less than $\Delta\xi_{min}$, we always retain the tagged point and remove its nearest neighbour. This procedure can significantly increase the lifetime of the tagged vortex-filament points. Nevertheless, these tagged points may still get removed in some situations. For instance, when two adjacent points are both tagged points with a separation less than $\Delta\xi_{min}$, one of them will be removed. Furthermore, if a tagged point exists in a vortex loop with a length less than $5\Delta\xi_{min}$, it will be removed together with the loop. Finally, due to the vortex reconnection procedures~\cite{Tsubota-2000-PRB,Baggaley-2012-JLTP}, there is a certain chance that a tagged point may get removed when it is involved in a reconnection event. Whenever a tagged point is removed, we randomly choose a new point to tag so that there are always 200 tagged points at every time step. Nonetheless, due to the aforementioned loss mechanisms, the averaged lifetime of the tagged points is still limited to few seconds.

\begin{figure}[t]
\includegraphics[width=1.0\linewidth]{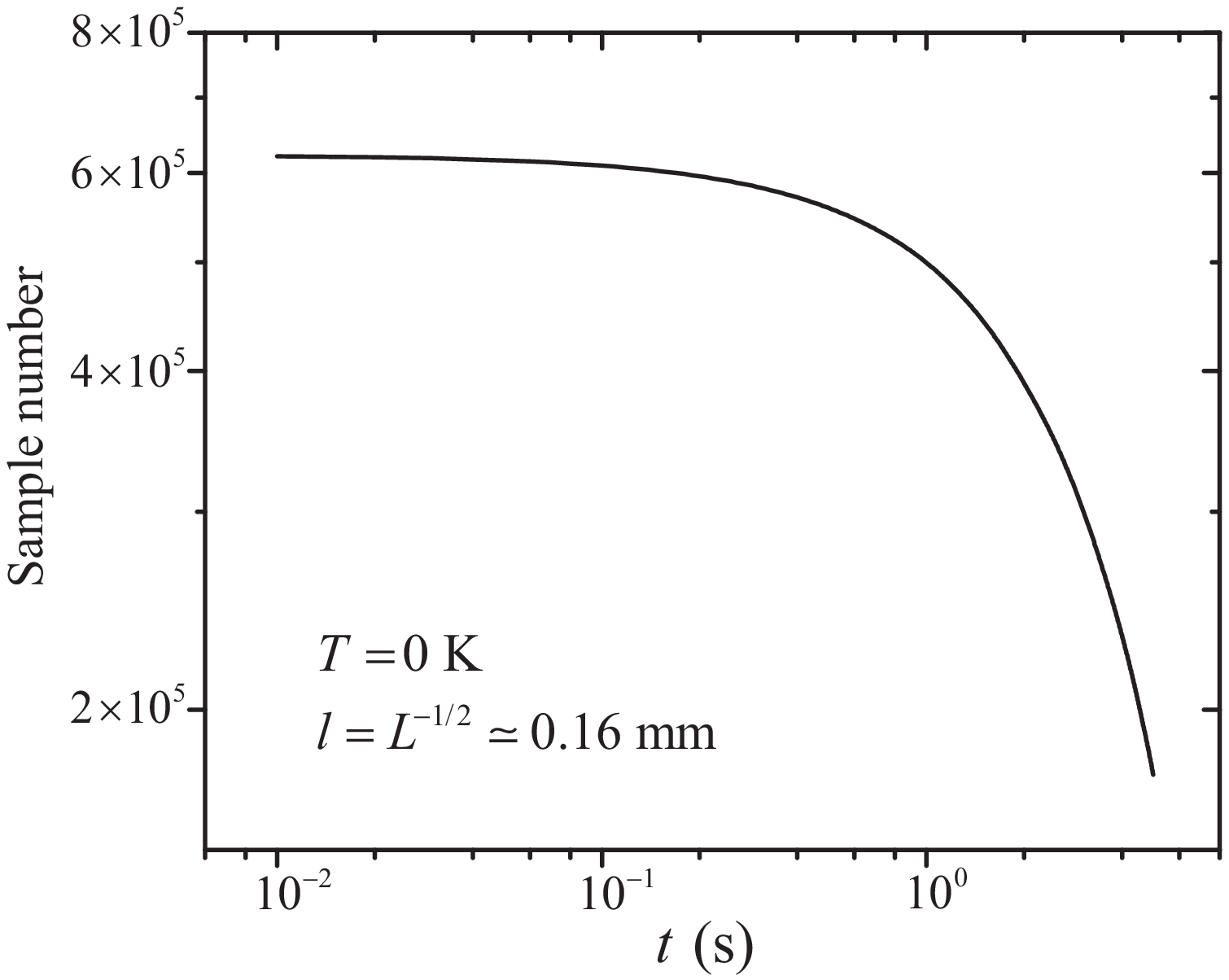}
\caption{Sample number versus diffusion time $t$ for a representative case at $T=0$~K with $\ell=0.16$~mm.}
\label{Suppl-Fig1}
\end{figure}

When we calculate the mean square displacement (MSD) $\langle\Delta x^2(t)\rangle=\langle[x(t_0+t)-x(t_0)]^2\rangle$, the number of vortex trajectories for ensemble averaging drops with increasing the diffusion time $t$. Fig.~\ref{Suppl-Fig1} shows a representative curve of the sample number versus $t$. Clearly, the sample number drops drastically as $t$ becomes greater than 1~s, which limits the range of the normal diffusion regime and increases the uncertainty of the fitted exponent $\gamma_2$. Another factor that also limits the accuracy of $\gamma_2$ is our computational box size. Note that our box size (i.e., 1~mm) is not too much larger than the typical line spacing $\ell$. Therefore, the vortex tangle is non-uniform at large scales even for tangles produced by random ring injections. Some large-scale variations of the vortex tangle can be clearly seen in Fig.~1~(a) in the main paper. This non-uniformity can lead to sizeable variations of $\gamma_2$ in the three axial directions.

\section{Vortex reconnection and displacement distribution}
For systems involving random walkers, it has been known that the superdiffusion can emerge as a consequence of the so-called L\'{e}vy flights, i.e., long-distance hops of the walkers~\cite{Zaburdaev-2015-RMP}. In a short time interval $\Delta t$, if one measures the displacement $\Delta x=x(t_0+\Delta t)-x(t_0)$ of the random walkers and calculates the displacement distribution $P(\Delta x,\Delta t)$, this distribution usually exhibits power-law tails $P(\Delta x,\Delta t)\propto|\Delta x|^{-\alpha}$ with $\alpha<3$ as caused by the L\'{e}vy events~\cite{Zaburdaev-2015-RMP}. After many steps, the resulted displacement distribution $P(\Delta x,t)$ can converge to a standard L\'{e}vy distribution with the same power-law tails~\cite{Bouchaud-1990-PR}. Mathematically, these flat tails would cause the MSD $\langle \Delta x^2\rangle$ of the walkers to diverge. However, one may introduce a pseudo MSD through a scaling argument and derive that $\langle \Delta x^2(t)\rangle\propto~t^{\gamma}$ with $\gamma$=$\frac{2}{\alpha-1}$~\cite{Bouchaud-1990-PR}. Therefore, an apparent superdiffusion of the walkers (i.e., $\gamma>1$) can emerge when $\alpha<3$. On the other hand, if the tails of $P(\Delta x,\Delta t)$ is not sufficiently flat (i.e., if $\alpha\geq3$), the central limit theorem then warrants a Gaussian distribution of $P(\Delta x,t)$, which would then lead to a normal diffusion of the walkers~\cite{Bouchaud-1990-PR}.

\begin{figure}[t]
\includegraphics[width=1.0\linewidth]{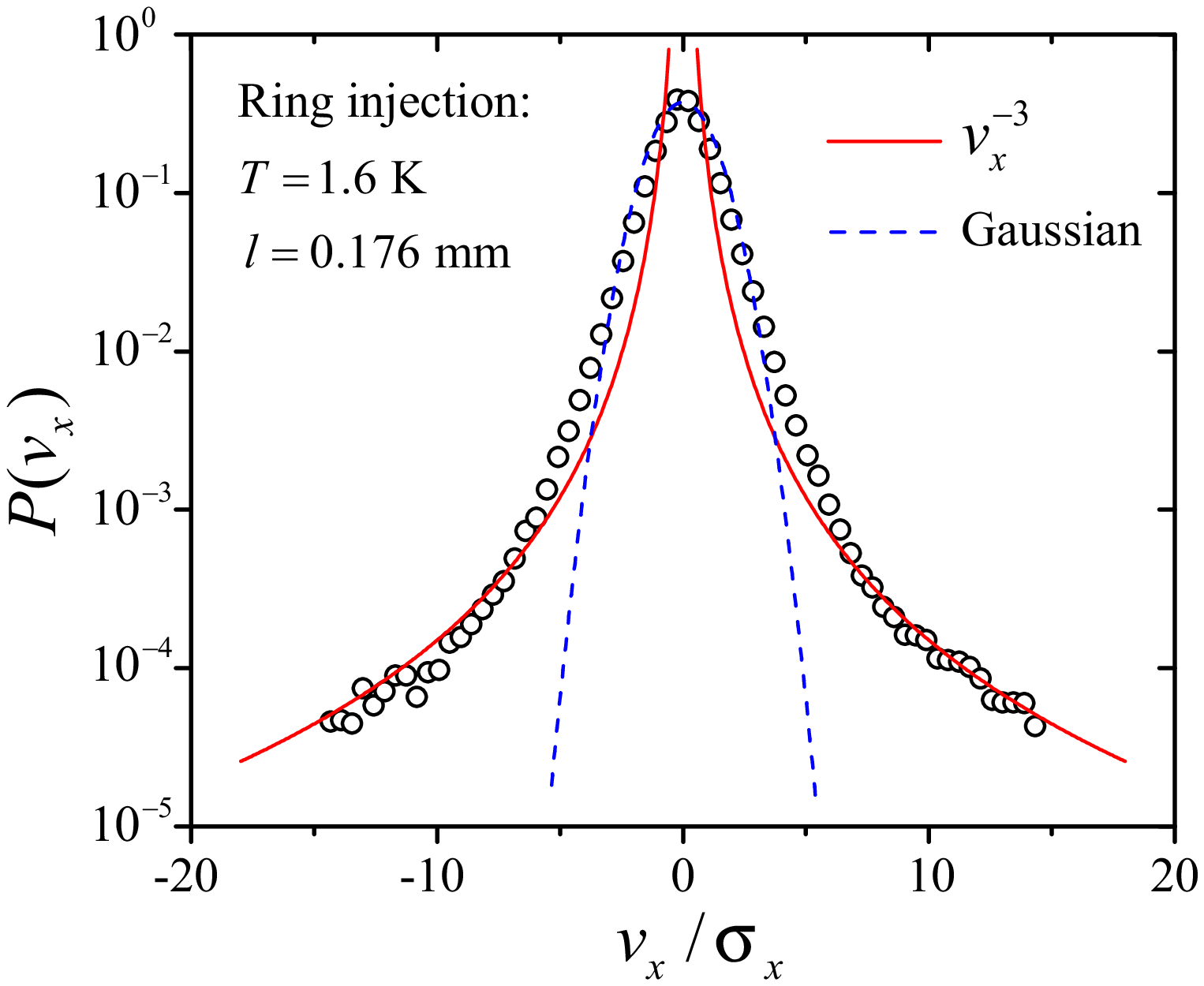}
\caption{A representative calculated probability density function $P(v_x)$ of the vortex velocity $v_x$ in the $x$-direction. The analysis was performed for the random vortex tangle produced by ring injections at 1.6~K with $\ell=0.176$~mm.
} \label{Suppl-Fig2}
\end{figure}

For quantized vortices in a random vortex tangle in He II, the vortex-filament points move chaotically as controlled by both the local superfluid velocity induced by all the vortices and the mutual friction force from the normal-fluid component~\cite{Donnelly-1991-B}. Interestingly, the vortex-filament points do exhibit occasional long-distance hops over short time intervals when they are sufficiently close to other vortex filaments or are involved in vortex reconnections~\cite{Paoletti-2008-PRL}. This is because when the vortex-filament points are close to the cores of other vortices, the induced velocity according to the Biot-Savart law can become exceptionally large~\cite{Donnelly-1991-B}. These high vortex-velocity occurrences are known to lead to non-Gaussian $|v|^{-3}$ tails of the vortex-velocity distribution~\cite{Paoletti-2008-PRL,Mastracci-2019-PRF}. In Fig.~\ref{Suppl-Fig2}, we show the calculated probability density function (PDF) $P(v_x)$ of the vortex velocity $v_x$ in the $x$-direction for a representative tangle produced by ring injections at 1.6~K with $\ell=0.176$~mm. This PDF is generated by analyzing the velocities of all the tracked vortex-filament points in the steady-state time window. The $|v_x|^{-3}$ tails are clearly visible. Similar results were reported for vortices in a drop of Bose-Einstein Condensate~\cite{White-2010-PRL} and in He II counterflow turbulence~\cite{Adachi-2011-PRB}. Since the displacement of a vortex-filament point over a short time interval $\Delta t$ is $\Delta x=v_x\cdot\Delta t$, the displacement distribution of the vortex-filament points should acquire similar power-law tails $P(\Delta x,\Delta t)\propto|\Delta x|^{-3}$.

\begin{figure}[t]
\includegraphics[width=1.0\linewidth]{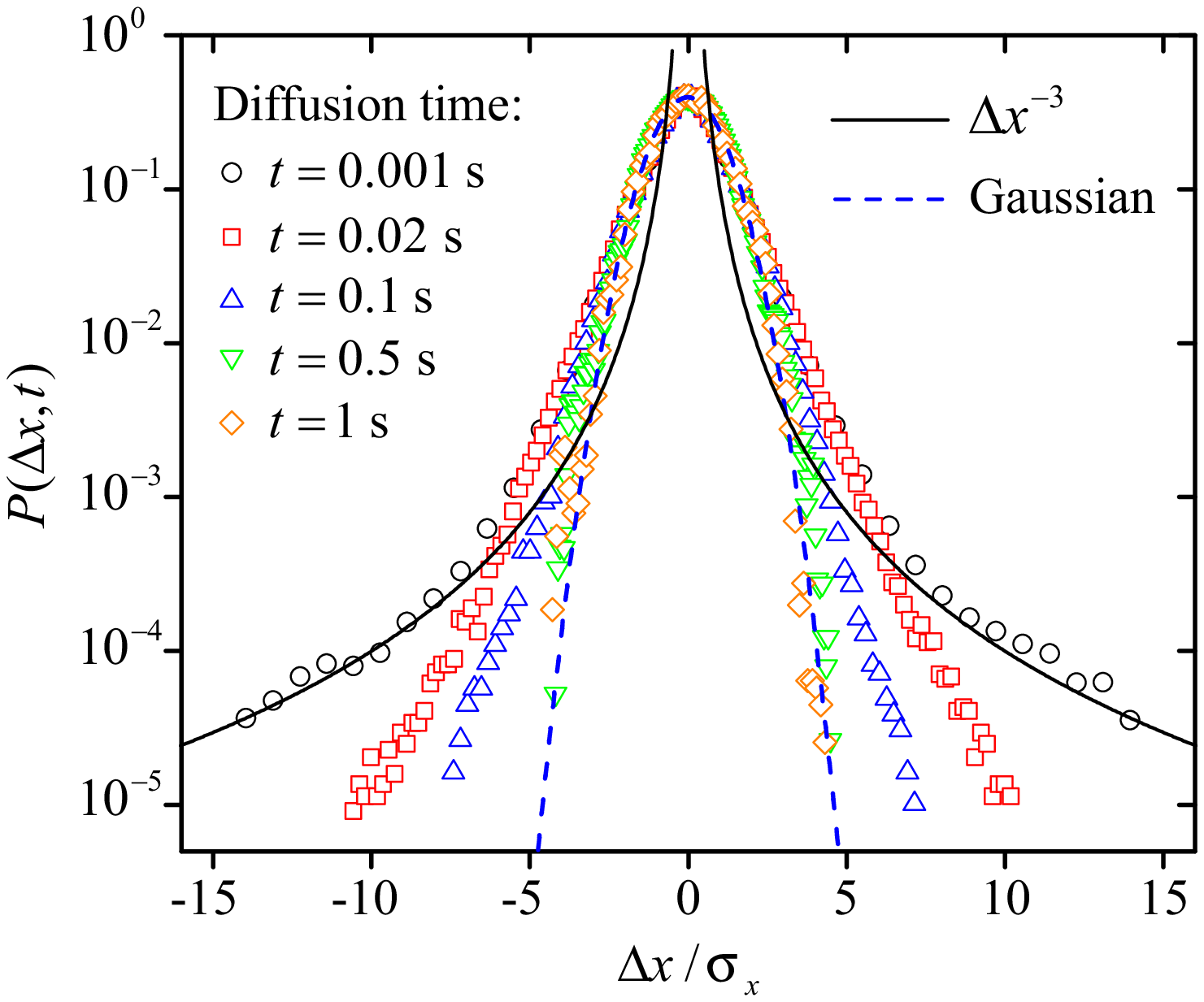}
\caption{Vortex-point displacement distribution $P(\Delta x,t)$ at various diffusion time $t$. The analysis was performed for the same vortex-tangle case as in Fig.~\ref{Suppl-Fig2}.
} \label{Suppl-Fig3}
\end{figure}

However, these flat tails of $P(\Delta x,\Delta t)$ do not lead to the superdiffusion of the vortices. This is because the velocities of the vortex-filament points drop rapidly as they move away from nearby vortex cores or the reconnection sites. Therefore, over longer time $t$ the total displacement of a vortex-filament point $\Delta{x}$=$\int_0^t{v_x(t')dt'}$ would rarely exhibit exceptionally large values. As a consequence, the tails of $P(\Delta x,t)$ are gradually suppressed as $t$ increases. To see this effect, we show in Fig.~\ref{Suppl-Fig3} the calculated displacement distribution $P(\Delta x,t)$ at various diffusion time $t$ for the same vortex-tangle case as presented in Fig.~\ref{Suppl-Fig2}. At the smallest diffusion time $t=10^{-3}$~s, $P(\Delta x,t)$ exhibits clear $|\Delta x|^{-3}$ power-law tails, which reflects the $|v_x|^{-3}$ tails of the vortex velocity distribution. But as $t$ increases to over 0.5~s, these power-law tails are suppressed to nearly a Gaussian form. This $P(\Delta x,t)$ behavior was also observed experimentally~\cite{Tang-2021-PNAS}. Therefore, despite the existence of some long-distance hops of the vortex-filament points at small time steps, their statistical weight is not sufficient to render the observed vortex superdiffusion. The true mechanism is the vortex velocity temporal correlation as discussed in the main text.

\section{Supplemental Movies}
\textbf{Movie S1:} Evolution of the vortex filaments in a random tangle produced by vortex-ring injections at 0 K with a steady-state mean vortex-line spacing $\ell=0.188$~mm. The red dots represent the tracked vortex-filament points for diffusion analysis.

\textbf{Movie S2:} Evolution of the vortex filaments in a tangle produced by counterflow at 1.6~K with a steady-state mean vortex-line spacing $\ell=0.205$~mm. The red dots represent the tracked vortex-filament points for diffusion analysis.

\bibliographystyle{naturemag}
\bibliography{Suppl-reference}